\documentclass[prx,amsmath,amssymb,reprint,superscriptaddress,nofootinbib]{revtex4-2}

\usepackage{braket}
\usepackage{bm}
\usepackage{booktabs}
\usepackage{tabularx}
\newcolumntype{L}{>{\raggedright\arraybackslash}X}
\usepackage{graphicx}
\usepackage{float}
\usepackage{multirow}
\usepackage{algorithm}
\usepackage{algpseudocode}
\algrenewcommand\textproc{}

\begin{document}

\title{Quantum-Computing Self-Consistent Kohn--Sham DFT:\\
       Plane-Wave-Orthonormalized Orbitals with a Dual-Basis Quantum Eigensolver}

\author{Lazaro Calderin}
\email{Contact author: calderin@ionq.com}
\affiliation{IonQ Inc., College Park, Maryland 20740, USA}

\begin{abstract}
We present a first-quantization, all-electron, full-potential Kohn--Sham
density functional theory method for quantum computers, based on a
classically optimized quantum-circuit eigensolver for the self-consistent
field (SCF) eigenproblems in any orthonormal basis.  The plane-wave-orthonormalized-orbital (PWOO) basis
combines plane waves with atomic orbitals (AOs) orthonormalized against
them and one another.  The eigensolver has a dual-basis
form: after full-basis bootstrap steps, it solves each SCF step in a
contracted basis of $K$ functions ($\lceil\log_2K\rceil$ qubits), updated
in the full basis when the residual outside it exceeds a threshold.  For
an H$_8$ chain (SVWN, $K=8$), the state-vector-emulated SCF agrees with
classical calculations within $0.02$\,mHa for $32\,768$ plane waves
and PWOO bases up to cc-pVTZ; contraction changes these PWOO energies by at most $5.7\,\mu$Ha.  With emulated shot noise, the STO-3G PWOO SCF
converges near the noise-free fixed point.  On IonQ Forte-1,
three-qubit-register readout reproduces the noise-free PWOO--AO energy
difference within one standard deviation.
\end{abstract}

\maketitle

\section{Introduction}
\label{sec:intro}

Kohn--Sham density functional theory (KS-DFT)
\cite{Hohenberg1964,Kohn1965} is the most widely used method for
electronic-structure simulation on conventional computers in quantum
chemistry, materials science, and condensed-matter physics, and is
therefore the reference for quantum-computing electronic-structure
methods.  This baseline is not limited to
local or semilocal functionals: hybrid functionals, DFT+$U$, and DFT+dynamical mean-field theory (DMFT) extensions are widely used when semilocal functionals do not describe exchange or
correlation adequately
\cite{Becke1993,Anisimov1991,Kotliar2006}.

Most molecular quantum-computing electronic-structure formulations
map a fixed one-particle-basis Hamiltonian to qubits in second
quantization and use the variational quantum eigensolver (VQE), quantum phase
estimation (QPE), or related eigensolvers
\cite{McClean2016,Bauer2020,McArdle2020}.  Fault-tolerant first-quantized and plane-wave algorithms~\cite{Babbush2018,Babbush2019,Su2021,Berry2024,Georges2025,Ivanov2025} are a related but distinct line:
they seek the interacting many-electron wave function directly.
For DFT specifically, there have been proposals to compute one component
of the calculation on quantum hardware, for example an exchange-correlation input or a learned functional
\cite{Hatcher2019,Baker2020,Sheridan2024}, and compact one-particle
encodings~\cite{Cerasoli2020,Sherbert2021} and density-matrix
formulations~\cite{Ko2023} have been proposed for self-consistent
calculations.
Within self-consistent quantum-computing DFT, Senjean, Yalouz, and
Sauban\`ere~\cite{Senjean2022} provide the closest prior simulated
molecular demonstration.  Their formal mapping, written for a finite orthonormal KS basis, encodes the $N\times N$ KS matrix on $\log_2N$ qubits; their proof-of-concept simulations include
a lattice site-occupation DFT calculation for a one-dimensional
Hubbard model and a molecular KS-DFT calculation for an H$_8$ linear
chain with the SVWN local-density functional~\cite{Slater1951,Vosko1980} in a
L\"owdin-orthonormalized STO-3G atomic-orbital (AO) basis~\cite{Hehre1969}.  In the
H$_8$ case, subspace-search (ensemble) VQE~\cite{Nakanishi2019} solves for the occupied finite-basis KS orbitals
at each self-consistent iteration.

Here we present a first-quantization, all-electron, full-potential
formulation of quantum-computing (QC) KS-DFT.  It requires only the KS matrix in an
orthonormal finite basis and therefore applies in principle to any such
basis, including periodic ones; we give the $\mathbf k$-dependent form of
the plane-wave-orthonormalized-orbital (PWOO) basis for periodic systems (Sec.~\ref{sec:pwoo-main}) and
demonstrate it at $\mathbf k=0$ for an isolated molecule.  We do not claim a wall-time advantage over classical DFT.  The method has two properties.  First, a single plane-wave cutoff
enlarges the basis systematically until the total energy converges to a
chosen tolerance.  Second, a dual-basis contraction solves every
self-consistent-field (SCF) step after a few initial full-basis steps on
a fixed register, with full-basis solves only for subspace updates.  For
the pure plane-wave (PW) sizes for which we evaluated it ($n_q=6$--12),
the Pauli L1 norm of the contracted Hamiltonian, which sets its
per-evaluation shot bound, stays at $0.59$--$0.75$\,Ha, whereas that of
the full Hamiltonian grows from $7$ to $233$\,Ha.

The central component of the method is a quantum-circuit eigensolver
(the QC eigensolver) that is called at every step of the KS-DFT SCF loop.  Its
classical steps are the angle optimization, the orthonormalization of
the orbitals, and, in the block-parallel runs (below), a small
Rayleigh--Ritz step in the span of candidate states, which in the
state-vector block-parallel runs includes two residual directions
computed classically from the Hamiltonian matrix.  A single KS orbital is encoded on a qubit register
spanning the chosen finite basis, and the multi-orbital occupied
subspace is obtained orbital by orbital by
Rotosolve~\cite{Nakanishi2020,Ostaszewski2021,Wierichs2022},
penalty deflation~\cite{Higgott2019}, and in-loop modified
Gram--Schmidt orthonormalization (MGSO).  The outer loop reassembles $H_{\rm KS}[\rho]$ from the density produced by those orbitals,
mixes the density, evaluates the KS total energy or Mermin free
energy for the current density, and iterates to self-consistency.  The same eigensolver is used for pure PW, for the
PWOO basis introduced below, and for
their contracted subspaces.  In the full-basis solves of the largest pure-PW run and in the PWOO
calculations (apart from the serial comparison runs of
Table~\ref{tab:pwoo-contracted}), Rotosolve updates
are computed in parallel over disjoint blocks of angles
(block-parallel Rotosolve, a block-coordinate descent) and combined by a
Rayleigh--Ritz step~\cite{Parlett1998}.

The dual-basis contraction is formed from the $K$ lowest eigenstates
after a few initial SCF steps in the full basis (bootstrap steps; one to
five here), the basis in which the KS Hamiltonian is assembled (PW or PWOO; Sec.~\ref{sec:contraction}).  Whenever the residual norm outside the contracted subspace exceeds a threshold, the contracted basis is recomputed at the current density
with $K$ fixed, so the register does not grow.  Together, the contracted
solve, the residual test, and the subspace update form a dual-basis
eigensolver (the name refers to the two bases between which the
eigensolver alternates, not to the plane-wave dual basis of
Ref.~\cite{Babbush2018}): at each SCF step it returns approximations to the $K$ lowest eigenpairs of
$H_{\rm KS}[\rho]$ in the full basis, with the residual of the checked
states outside the contracted subspace below the threshold (all $K$ states
for PWOO; for pure PW the four lowest, so that at $n_q=6$ the fifth,
singly occupied state is not checked).

The pure-PW basis can be enlarged systematically until the energy
converges to a chosen tolerance, but for all-electron orbitals the
convergence is slow: even $N_{\rm PW}=32\,768$ ($n_q=15$, our largest
QC-emulated register) is far from converged; at that size a
state-vector-emulated QC SCF run takes more than an hour, and the
per-evaluation shot bound of the full Hamiltonian already reaches
$2\times10^{10}$ shots at $n_q=12$ (Sec.~\ref{sec:stage2}).
To reduce the large number of plane waves needed to describe cusps and
orbital oscillations near nuclei, we introduce PWOO, an orthonormal mixed basis that reaches a given energy
with far fewer functions, whose span is that of the retained PW block
together with AOs orthonormalized to that block.  The PWOO operator matrices are
assembled in non-orthogonal PW--PW, PW--AO, and AO--AO blocks of the raw
PW and AO functions and then transformed to the final orthonormal basis, so the
QC eigensolver again receives a Hamiltonian in an orthonormal
basis.  The shot-based PWOO runs use the same block-coordinate search; on
the full register the energy is estimated by measuring the blocks of
$H_{\rm PWOO}$ between plane waves and orthonormalized orbitals (OOs)
separately, and on the
contracted register by grouped Pauli measurement.

Sec.~\ref{sec:method} describes the encoding, eigensolver, SCF loop, and
dual-basis contraction, Sec.~\ref{sec:pwoo-main} the PWOO basis, and
Sec.~\ref{sec:stage2} the calculations: pure PW, PWOO, shot-based PWOO,
contracted PWOO, and the trapped-ion readout, followed by a discussion
of prior work and open problems.

The calculations reported here are all-electron and full-potential in
the operational sense used throughout: bare Coulomb nuclei, no
pseudopotential or effective core potential, and no
muffin-tin/interstitial partition.  The
linear H$_8$ chain (1.4\,bohr spacing) sits in a cubic cell of side
$L=20$\,bohr, and the Coulomb interactions use a spherically truncated Coulomb
kernel~\cite{Jarvis1997,Rozzi2006} with cutoff radius $R_c=L/2$, which removes periodic-image interactions for the isolated
molecule.  We use the spin-restricted SVWN local-density functional,
Slater exchange~\cite{Slater1951} with Vosko--Wilk--Nusair
correlation~\cite{Vosko1980}.  The shot results are emulated: on the full register each energy estimate
is the noise-free expectation value plus Gaussian noise with the variance
of the shot estimator, and on the contracted register the measurement
counts of grouped Pauli strings are sampled; the hardware experiment of
Sec.~\ref{sec:hardware} is a real-device readout of the converged
states, not a hardware SCF loop.

\section{Method}
\label{sec:method}
\label{sec:ks-on-qubits}
\label{sec:ansatz}
\label{sec:multi-orbital}
\label{sec:scf}
\label{sec:contraction}

\subsection{Qubit encoding, eigensolver, and SCF loop}
The qubit register represents one KS orbital at a time: a basis of
$N=2^{n_q}$ orthonormal functions is held on $n_q$ qubits, one function
per computational basis state.  When the basis size is not a power of
two, the unused states are suppressed by an energy penalty (padding);
every calculation reported here has $N=2^{n_q}$, so none is padded.  In a pure
PW basis, the computational basis states are identified with the reciprocal-lattice
vectors $\mathbf G$ of a cubic $n\times n\times n$ grid ($N=n^3$); in PWOO, they are identified with
the final orthonormal PWOO functions.  In either case an orbital is
prepared as
\begin{equation}
    |\psi_m\rangle = \sum_{j=0}^{N-1} A_{m,j}e^{i\varphi_{m,j}}|j\rangle,
    \qquad \sum_j A_{m,j}^2=1 ,
\end{equation}
using the M\"ott\"onen amplitude tree and phase layer \cite{Mottonen2005}.  The ansatz spans the full one-particle Hilbert
space of the chosen finite basis.  Each parameter is a single
rotation angle, so the energy along one coordinate is a low-order
trigonometric polynomial.  Phase coordinates are minimized from
three shifted evaluations, and amplitude coordinates from five
shifted evaluations followed by a one-dimensional classical
minimization.  This is a Rotosolve-type coordinate update
\cite{Nakanishi2020,Ostaszewski2021,Wierichs2022}.

For both state-vector and simulated-shot runs, the angles can be partitioned into disjoint blocks, formed from subtrees of the
M\"ott\"onen amplitude tree, and optimized in parallel
(block-parallel Rotosolve): the eight subtrees rooted at level $l=3$ of the tree form the blocks, and the seven amplitude angles above
them are updated serially in every round (on the three-qubit contracted
register, four one-angle blocks below three top angles).  Each block starts from the
same orbital, performs a local Rotosolve descent over its own amplitude
angles (updated four at a time from a common snapshot), and returns a
candidate vector.  Instead of accepting the lowest-energy candidate, we update the orbital by a Rayleigh--Ritz step~\cite{Parlett1998}, as in the locally optimal block preconditioned conjugate gradient
(LOBPCG) method~\cite{Knyazev2001}, in the span of the current state, the block candidates, and the
state obtained by applying all block updates together.  Two residual
directions are added: $(H-\epsilon)\psi$, with $\epsilon$ the Rayleigh
quotient of the current state, and its diagonally preconditioned
(Davidson~\cite{Davidson1975}) form, each computed with the undeflated
$H$ and then projected orthogonal to the lower eigenstates.  The
$n_q=15$ run also includes the states at the start of up to eight
preceding rounds of the same eigenstate solve.  The lowest
Ritz vector is converted back to angles, so in these runs the phases
change only through this step.  The residual directions are
computed classically from the Hamiltonian matrix and are therefore
omitted in the shot-based eigensolves of the SCF runs.  One round of the eigensolver
consists of the block searches and this Rayleigh--Ritz step.  This
parallel block-coordinate descent is the angle search of the QC eigensolver used in the $n_q=15$ pure-PW run
and in the PWOO calculations; the $n_q\le12$ pure-PW runs, the
contracted solves of the $n_q=15$ run, and the rows
marked serial in Table~\ref{tab:pwoo-contracted} use Gauss--Seidel
Rotosolve, which updates one angle at a time.  In the $n_q=15$
pure-PW run, the fixed-Hamiltonian shot test and the full-register
shot-SCF runs of Sec.~\ref{sec:stage5} (except the test runs without
randomized subsets), and the block-parallel
STO-3G run of Table~\ref{tab:pwoo-contracted}, each round is made cheaper by randomized subsets: the
angles of every block are split into four interleaved subsets, and in a
given round each block updates only one subset, drawn at random from a
generator with a fixed seed, so that the angles are covered over
successive rounds.  In these STO-3G runs (Tables~\ref{tab:pwoo-shot-scf-gate}
and~\ref{tab:pwoo-contracted}), each eigenstate solve uses
a fixed schedule of 128 rounds; all other solves stop when the energy
drop per round falls below a tolerance, after at least 16 rounds in the
block-parallel PWOO solves.

Multi-orbital KS eigensolving is done by penalty deflation~\cite{Higgott2019}.  After orbitals $0,\ldots,m-1$ are found, orbital $m$ minimizes $\langle\psi|H^{(m)}|\psi\rangle$ with
\begin{equation}
    H^{(m)} = H_{\rm KS}
      + \mu\sum_{j<m}|\psi_j\rangle\langle\psi_j| .
\end{equation}
The penalty $\mu$ is the spectral norm $\|H_{\rm KS}\|_2$, estimated by
power iteration, so that the penalized previously found orbitals are shifted upward relative to the target state.  Because the penalty is quadratic in the overlap, its gradient vanishes as
the overlap goes to zero, so a finite number of Rotosolve rounds leaves
nonzero overlaps between near-degenerate orbitals.  We therefore orthonormalize each newly returned
state against the previously found orbitals with MGSO, re-evaluate its
energy on the orthonormalized state, and use that state in the penalty for the next orbital.  With orthonormal orbitals, the density
$\rho(\mathbf r)=\sum_m f_m|\psi_m(\mathbf r)|^2$ is invariant under unitary
rotations within subspaces of equal occupation.

The self-consistent loop is otherwise conventional.  At iteration
$t$, the classical side assembles $H_{\rm KS}[\rho^{(t)}]$, the QC
eigensolver returns orbitals $\psi_m$ and eigenvalues $\epsilon_m$, occupations
$f_m\in[0,2]$ (spin-restricted) are assigned either by an explicit fixed (pinned)
pattern ($[2,2,2,1,1,0,0,0]$ at $n_q=6$, which half-fills the near-degenerate highest occupied pair) or by Fermi--Dirac (FD) smearing, and the density is mixed for the next Hamiltonian, either
linearly with mixing parameter $\alpha$ or by Anderson
mixing~\cite{Anderson1965}, which extrapolates from a short history of
previous input and output densities.  Total energies are reported as KS total energies $E_{\rm KS}$, including
the nuclear repulsion energy $V_{\rm NN}$; at finite smearing the
convergence test is applied to the Mermin free energy
$F=E_{\rm KS}-TS$~\cite{Mermin1965}, with $S$ the entropy of the
fractional occupations and $T$ the smearing temperature; all energies
reported in this paper are $E_{\rm KS}$.

The main runs use three fixed-tolerance convergence criteria; the other
tests are given in Appendix~\ref{app:details}.  In the
\emph{pure-PW criterion}, the runs stop when $|\delta F|<10^{-5}$\,Ha for the change between SCF steps
and the largest change of the output density between successive steps is below $10^{-4}$\,bohr$^{-3}$.  In the \emph{three-step PWOO criterion}, used by the noise-free PWOO runs of Tables~\ref{tab:pwoo-main} and~\ref{tab:pwoo-contracted}, they stop when, on three consecutive SCF steps, $|\delta E|<10^{-6}$\,Ha and the change of the occupied
projector $P_{\rm occ}=\sum_{m\in{\rm occ}}|\psi_m\rangle\langle\psi_m|$ between
steps satisfies $\|\Delta P_{\rm occ}\|_F<10^{-4}$ (Frobenius norm), and the
energy varies by less than $10^{-6}$\,Ha over those steps and the step before them; the QC and
classical PWOO runs use the same criteria.  The shot-based runs use the \emph{shot-run criterion}, the looser
single-step tolerances $|\delta E|<10^{-5}$\,Ha and
$\|\Delta P_{\rm occ}\|_F<3\times10^{-4}$.

Because the M\"ott\"onen
parametrization gives the orbital amplitudes explicitly as functions of
the angles, the classical side obtains the orbitals, the density, and the
Rayleigh--Ritz combinations from the angles found by the QC eigensolver,
without state tomography; errors in the optimized angles therefore
propagate directly into these quantities, and the shot-SCF results
include them.

To reduce the register width of the eigensolves repeated at each SCF
step, we use a dual-basis contraction.  As noted in the Introduction, the
name is unrelated to the
plane-wave dual basis of Babbush et al.~\cite{Babbush2018} and to dual-basis
DFT methods~\cite{Liang2004}, and ``contracted'' refers to this subspace,
not to contracted Gaussian functions.
We call the basis in which $H_{\rm KS}$ is assembled (pure PW or PWOO)
the full basis, and its register the full register.  A
bootstrap solve in the full basis builds a $K$-column
orthonormal matrix $U$ from the $K$ lowest orbitals returned by the QC eigensolver ($K=8$ in
all QC calculations here); its columns span the contracted basis.
Subsequent SCF steps solve
\begin{equation}
    \tilde H = U^\dagger H_{\rm KS} U
\end{equation}
on $n_q'=\lceil\log_2K\rceil$ qubits, the contracted register.  Each
contracted eigenstate $\tilde\psi_m$ is lifted back to the full basis,
$\psi_m=U\tilde\psi_m$, and its residual norm outside the contracted subspace
\begin{equation}
    \eta_m=\|(1-UU^\dagger)(H_{\rm KS}-\epsilon_m)\psi_m\|
\end{equation}
measures the part of its eigenvalue residual that lies outside the
contracted basis.  When $\max_m\eta_m$ (Euclidean norm, in Ha) exceeds a threshold
($2.5\times10^{-3}$\,Ha in all runs of Tables~\ref{tab:main-results} and~\ref{tab:pwoo-contracted}), $U$ is recomputed within the same SCF step, at the same density: the QC eigensolver solves
the full-basis problem, warm-started from the lifted states,
and its $K$ lowest eigenstates become the new columns of $U$.  This differs from Davidson subspace growth~\cite{Davidson1975}: the register width stays
fixed, while the subspace basis is updated.

\subsection{PWOO basis}
\label{sec:pwoo-main}

\begin{figure*}[t]
\centering
\includegraphics[width=0.95\textwidth]{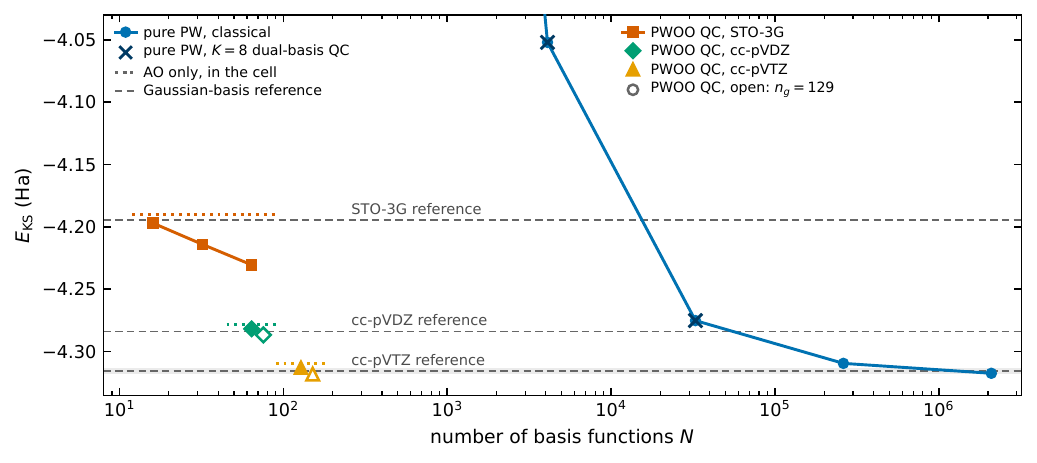}
\caption{H$_8$/SVWN energy versus basis-set size.  Dashed lines:
free-space Gaussian-basis references computed with PySCF; dotted lines: AO-only references in the same
cell.  Circles: classical PW reference, enlarged with the single cutoff
parameter; crosses: $K\!=\!8$ dual-basis QC energies
(Table~\ref{tab:main-results}; the $n_q=6$ and 9 points lie above the
plotted range, and the circles at $N=2.6\times10^5$ and $2.1\times10^6$
are classical calculations only).  Squares: STO-3G PWOO energies; every
PWOO point lies below its AO-only reference, as the nested-basis
variational bound requires.  cc-pVDZ and cc-pVTZ PWOO points are shown
for $n_g=65$ (filled symbols) and $n_g=129$ (open symbols, shifted right
for visibility); shaded: $\pm1.6$\,mHa around the cc-pVTZ reference.  The PWOO points are QC SCF runs on the full register;
the contracted $K\!=\!8$ runs of Table~\ref{tab:pwoo-contracted}
(STO-3G with $N_{\rm PW}=8$, and cc-pVTZ) differ from the corresponding
points by at most $35.1\,\mu$Ha and are not shown.  Reference values and provenance in
Table~\ref{tab:pwoo-main} and Appendix~\ref{app:details}.}
\label{fig:basis-conv}
\end{figure*}

A plane-wave basis can be enlarged systematically until the energy
converges to a chosen tolerance, but the nuclear cusps of all-electron
orbitals make the required number of plane waves very large
(Fig.~\ref{fig:basis-conv}).  PWOO is also orthonormal, but it represents the short-wavelength,
near-nucleus part of the orbitals with AOs orthonormalized to a small
retained PW block instead of with high-$|\mathbf G|$ plane waves.
Let $\mathbf k$ be a Bloch vector of the first Brillouin zone.  The
retained PW set $\mathcal P_{\mathbf k}$ consists of the $N_{\rm PW}$ plane
waves $|\mathbf k+\mathbf G\rangle=\Omega^{-1/2}e^{i(\mathbf k+\mathbf G)\cdot\mathbf r}$
of lowest kinetic energy $\frac12|\mathbf k+\mathbf G|^2$, with $\Omega$ the
cell volume, $N_{\rm PW}=2^{n_q}-N_{\rm AO}$ so that the basis fills the
register, and the plane waves of the last, partially filled shell taken
in a fixed index order.  A kinetic-energy cutoff that keeps complete
shells, with any remaining register states padded, is the general
alternative and is supported by the implementation.  The AO basis
$\mathcal Q=\{q_a\}$ enters as Bloch sums
$\chi_{a,\mathbf k}(\mathbf r)=\sum_{\mathbf R}e^{i\mathbf k\cdot\mathbf R}q_a(\mathbf r-\mathbf R)$
over the lattice vectors $\mathbf R$; their overlaps with the PW block,
$\langle\mathbf k+\mathbf G|\chi_{a,\mathbf k}\rangle=\Omega^{-1/2}\hat q_a(\mathbf k+\mathbf G)$,
follow from the Fourier transform $\hat q_a$ of $q_a$.
With $\Pi_{\mathcal P_{\mathbf k}}$ the projector onto
${\rm span}(\mathcal P_{\mathbf k})$, the orthonormalized-orbital (OO)
block is
\begin{equation}
  \mathcal O_{\mathbf k} =
  {\rm orth}\{(1-\Pi_{\mathcal P_{\mathbf k}})|\chi_{a,\mathbf k}\rangle\}_{a},
  \qquad
  \mathcal B_{\rm PWOO}(\mathbf k)=\mathcal P_{\mathbf k}\cup\mathcal O_{\mathbf k} .
  \label{eq:oo-block}
\end{equation}
With $N_{\rm PW}=|\mathcal P_{\mathbf k}|$ and $N_O=|\mathcal O_{\mathbf k}|$
($N_O=N_{\rm AO}$ in all calculations reported here) and ${\rm orth}$
denoting canonical orthonormalization~\cite{Lowdin1970}, $N_{\rm PW}=0$
gives the AO-only limit and $N_O=0$ gives pure PW.  The PWOO
basis is orthonormal, a mixed basis of the kind introduced for plane waves and localized orbitals in
solids~\cite{Louie1979}.  In orthogonalized plane waves~\cite{Herring1940}
the plane waves are orthogonalized to core states; here the AOs are
orthogonalized to the plane waves.  Gaussian and plane-wave functions are
also combined in the Gaussian and plane-wave (GPW) and Gaussian and
augmented-plane-wave (GAPW) methods~\cite{Lippert1997,Lippert1999}.
This construction gives a direct numerical check, the nested-basis
variational bound.  Because the AO subspace is contained in ${\rm span}(\mathcal B_{\rm PWOO})$, the PWOO
variational minimum, computed with the same Hamiltonian, boundary
condition, occupation rule, and exchange--correlation treatment, must be
no higher than the AO-only minimum.  A violation would indicate an error in an operator block or unit-cell
integral, a mismatch in settings between the two calculations, or
convergence of one SCF to a different solution; satisfying the bound is
necessary but not sufficient for correctness.  Here finite-cell quantities are computed in the $L=20$\,bohr cell
with the unit-cell integration described below, the AO-only reference
is the finite-cell calculation in the AO basis alone ($N_{\rm PW}=0$),
and free-space Gaussian-basis references are standard molecular
calculations with no cell.

Operator assembly is done in non-orthogonal (raw) $P$--$P$, $P$--$Q$, and
$Q$--$Q$ blocks at each $\mathbf k$ and then transformed to the PWOO basis:
\begin{equation}
\begin{aligned}
H_{\rm raw}(\mathbf k) &=
\begin{pmatrix}
H_{PP} & H_{PQ}\\
H_{QP} & H_{QQ}
\end{pmatrix},\\
H_{\rm PWOO}(\mathbf k) &= C_{\mathbf k}^\dagger H_{\rm raw}(\mathbf k)\,C_{\mathbf k} ,
\end{aligned}
\label{eq:hpwoo}
\end{equation}
where $C_{\mathbf k}$ is the matrix of expansion coefficients of the PWOO
functions in the raw PW and Bloch-summed AO functions.  Unlike
augmented-plane-wave (APW) methods~\cite{Slater1937}, PWOO needs no
partition into muffin-tin spheres and an interstitial region: the overlaps and
the matrix elements are integrals over the unit cell (at $\Gamma$, the
AO--AO nuclear-attraction and Coulomb blocks are the free-space Gaussian
integrals), and the QC eigensolver receives only $H_{\rm PWOO}(\mathbf k)$.  The kinetic matrix is analytic
in all blocks: $\frac12|\mathbf k+\mathbf G|^2$ on the PW--PW diagonal,
$\frac12|\mathbf k+\mathbf G|^2$ times the PW--AO overlaps, and lattice
sums $\sum_{\mathbf R}e^{i\mathbf k\cdot\mathbf R}\langle q_a|\hat T|q_b(\cdot-\mathbf R)\rangle$
of Gaussian integrals between AOs.  The local potentials (electron--nuclear,
Hartree, exchange--correlation) are lattice-periodic, so their PW--PW
blocks depend only on $\mathbf G-\mathbf G'$ and are the same at every
$\mathbf k$; their blocks involving AOs carry the Bloch phases.  They are
functionals of the density
$\rho(\mathbf r)=\sum_{\mathbf k}w_{\mathbf k}\sum_m f_{m\mathbf k}|\psi_{m\mathbf k}(\mathbf r)|^2$,
with Brillouin-zone weights $w_{\mathbf k}$, so the KS problems at
different $\mathbf k$ are solved separately at each SCF step and coupled
only through $\rho$.  For a crystal, the truncated kernel is replaced by the periodic Coulomb
kernel.

For an isolated molecule in a cell large enough that the AOs do not
overlap their periodic images, the $\Gamma$ point $\mathbf k=0$ with
$w_{\mathbf 0}=1$ is sufficient, and we approximate the Bloch sums by the
AOs restricted to the cell, neglecting the tails of the periodic images
that re-enter the cell; the overlap and kinetic integrals then become
integrals of the AOs over the cell, and the kinetic energies of the plane waves become
$\frac12|\mathbf G|^2$.  All calculations reported here are of this
type, the implementation is restricted to $\Gamma$, and in the rest of
the paper we drop the subscript $\mathbf k$.  At $\Gamma$, the
electron--nuclear PW--PW block is evaluated in reciprocal space with the
truncated Coulomb kernel and the AO--AO block from analytic Gaussian
integrals; only the PW--AO block is integrated on the unit-cell grid
(Simpson rule, $n_g$ points per axis; $n_g=65$ unless stated otherwise).  The
Hartree matrix is built from the analytic unit-cell Fourier coefficients
of the PW and AO products, combined with the truncated Coulomb kernel
and extrapolated between two Fourier cutoffs;
the AO--AO block generated by the AO--AO part of the density is replaced by the analytic Gaussian Coulomb matrix.  The exchange--correlation potential
and energy are evaluated on the unit-cell grid; the PW--PW block is formed
from the Fourier coefficients of $v_{\rm xc}$ on that grid, and the blocks
involving AOs by quadrature on atom-centered grids (129 radial Simpson
points, 110 angular points).

\section{Results}
\label{sec:stage2}

All SCF results in this section start from a fixed initial guess, a
superposition of atom-centered Gaussian densities for pure PW and the
core-Hamiltonian (kinetic plus electron--nuclear) guess for PWOO, whose
eigenvectors also start the first PWOO eigensolve; no run is
warm-started from a previous solution.  All wall times are
emulation or classical computing times on eight cores of a desktop workstation
(Intel Core i9-14900K; the $n_q=6$ calculation of
Table~\ref{tab:main-results} and the PWOO runs of Tables~\ref{tab:pwoo-main}
and~\ref{tab:pwoo-contracted} on an Intel Core Ultra 9 285K): they
compare algorithmic variants within this work and are not
classical-DFT or hardware cost claims.
Table~\ref{tab:main-results} gives the pure-PW results for a sequence of register sizes.  At
fixed $K=8$ contracted dimension, the QC SCF energies stay within 0.14\,mHa of the classical PW SCF energies in the same basis from the smallest register to $n_q=15$.  At $n_q=6$
the QC and classical runs reach slightly different pinned SCF solutions
(highest occupied pair split by $4.1$\,mHa and $0.01$\,mHa,
respectively), so part of the $0.136$\,mHa difference there is between
nearly degenerate SCF solutions.  The main pure-PW result is the
$n_q=15$ calculation, with $N_{\rm PW}=32\,768$, a three-qubit contracted
solve, and $0.015$\,mHa agreement with the classical PW reference under the
residual threshold of $2.5\times10^{-3}$\,Ha.  At $n_q=9$, tightening the
threshold to $10^{-3}$\,Ha lowers the QC energy by 0.045\,mHa.  With the same $K$ and threshold (the classical $n_q=15$ run uses 3
bootstrap steps and 2 subspace updates), replacing the QC eigensolver by
a classical one changes the energy by $\le0.02$\,mHa at
$n_q\in\{9,12\}$ and $0.001$\,mHa at $n_q=15$; the differences in
Table~\ref{tab:main-results} are instead against the uncontracted
classical PW SCF.  A classical-reference
$K$-sweep at $n_q=9$ ($K\in\{4,8,16,24,32\}$) lies within $0.03$\,mHa of the
PW reference in $E_{\rm KS}$ ($0.021$, $0.028$, $0.015$, $0.012$, and
$0.010$\,mHa); the offset decreases from $K=8$ to $K=32$.

\begin{table*}[t]
\caption{Pure-PW results: dual-basis KS-DFT SCF on
H$_{8}$/SVWN at $K\!=\!8$ (contracted register $n_q'\!=\!3$),
state-vector emulation without shots.  Every calculation runs the QC
eigensolver (M\"ott\"onen ansatz, Rotosolve, penalty deflation) for the PW bootstrap,
subspace-update, and contracted solves, with residual threshold
$2.5\times10^{-3}$\,Ha; the $n_q\!=\!15$ calculation uses parallel block-coordinate descent with the
Rayleigh--Ritz step in its full-register solves.  Anderson mixing
with $\alpha\!=\!0.1$, history 6 (pinned) or $\alpha\!=\!0.3$,
history 2 (Fermi--Dirac, FD, smearing).  Steps: SCF steps, of which 1, 1, 1, and 3 use only contracted solves;
$n_{\rm upd}$: subspace updates;
$|\Delta E_{\rm KS}|=|E_{\rm KS}-E_{\rm KS}^{\rm PW}|$.
$E_{\rm KS}^{\rm PW}$ is the uncontracted classical Lanczos~\cite{Lanczos1950} PW SCF at the same
$n_q$.}
\label{tab:main-results}
\footnotesize
\begin{tabular}{@{}r r l r r r r r r@{}}
\toprule
$n_q$ & $N_{\rm PW}$ & Occupations & steps & $n_{\rm upd}$ &
$E_{\rm KS}$ (Ha) & $E_{\rm KS}^{\rm PW}$ (Ha) &
$|\Delta E_{\rm KS}|$ (mHa) & Wall \\
\midrule
$6$  & $64$      & pinned, $T\!=\!0$              & $7$ & $4$ & $+0.176241$ & $+0.176105$ & $0.136$ & $7$ s \\
$9$  & $512$     & FD, $k_BT\!=\!3$\,mHa          & $5$ & $1$ & $-2.807527$ & $-2.807574$ & $0.047$ & $1.9$ min \\
$12$ & $4096$    & FD, $k_BT\!=\!3$\,mHa          & $6$ & $2$ & $-4.051925$ & $-4.051947$ & $0.022$ & 1 h 59 min \\
$15$ & $32\,768$ & FD, $k_BT\!=\!3$\,mHa          & $8$ & $0$ & $-4.275129$ & $-4.275144$ & $0.015$ & 1 h 13 min \\
\bottomrule
\end{tabular}
\end{table*}

The measurement cost of the full and contracted bases can be compared
through the Pauli L1 norm.
We compare both at the same converged density (classical $K=8$ runs with
Fermi--Dirac occupations at $k_BT=3$\,mHa).
The L1 norm of the Pauli decomposition of the full PW Hamiltonian,
$\|h\|_1=\sum_j|h_j|$ for $H=\sum_j h_j\Sigma_j$ over Pauli strings
$\Sigma_j$, grows rapidly with the full register: $\|h\|_1=6.97$\,Ha at $n_q=6$, $32.0$\,Ha at
$n_q=9$, and $233$\,Ha at $n_q=12$.  The corresponding shot bound $(\|h\|_1/\delta)^2$ for one
energy evaluation to precision $\delta$, with shots sampled in proportion to $|h_j|$ (importance
sampling~\cite{Wecker2015}), is already $1.9\times10^7$,
$4.0\times10^{8}$, and $2.1\times10^{10}$ shots at chemical accuracy
($\delta=1.6$\,mHa).  The contracted Hamiltonian $\tilde H$ at $K=8$,
which is diagonal at this density, has L1 norms of order one Hartree
($0.75$, $0.59$, and $0.65$\,Ha at the same three basis sizes;
at $n_q=9$ the bound is $\sim\!1.4\times10^{5}$ shots per
evaluation), so the per-evaluation bound drops by a factor that grows
with the full-basis size, from $\sim\!9\times10^{1}$ at $n_q=6$ to
$\sim\!3\times10^{3}$ at $n_q=9$ and $\sim\!1\times10^{5}$ at
$n_q=12$, while the full-basis energy continues to decrease with $n_q$.
During the SCF, $\tilde H$ is not diagonal, and the subspace updates still
use the full register.
PWOO reduces the other cost, the full-basis dimension: the AOs represent
the orbitals near the nuclei, which reduces the number of plane waves
needed for a given energy.

PWOO reaches energies below the STO-3G reference with 16 basis
functions, whereas pure PW needs more than 4096 plane waves.  Table~\ref{tab:pwoo-main} shows
PWOO runs on their full registers ($N_{\rm PWOO}=2^{n_q}$, so no
padding is needed), formed from retained PWs plus
AOs orthonormalized to the PW block.  Each calculation gives an energy and a test of the nested-basis variational bound.  The STO-3G
sequence passes the test: with the same minimal AO basis (STO-3G), adding
8, 24, and 56 PWs lowers the AO-only energy by 7.07,
24.06, and 40.56 mHa, respectively, and all three rows are below the
free-space STO-3G Gaussian-basis reference (STO-3G is the AO basis of the
H$_8$ calculations of Senjean et al.~\cite{Senjean2022}).  The larger cc-pVDZ and cc-pVTZ~\cite{Dunning1989} rows satisfy the bound with margins of 3.47 and 3.65 mHa, but at $n_g=65$
they lie 2.3 and 2.6\,mHa above their free-space Gaussian-basis references.
This offset comes from the unit-cell grid.  At $n_g=65$ (spacing
$0.31$\,bohr) the AO-only energies lie 4.5 (STO-3G), 5.8 (cc-pVDZ), and
6.2\,mHa (cc-pVTZ) above the free-space references, which use the same AO
bases, and the plane waves lower the cc-pVDZ and cc-pVTZ energies by only
about 3.5\,mHa.  The grid error decreases more slowly with $n_g$ for the
correlation-consistent bases, whose tightest hydrogen s exponents ($13.0$ and
$33.9$\,bohr$^{-2}$, against $3.4$\,bohr$^{-2}$ for STO-3G) are less well
resolved.  With $n_g=129$ (spacing $0.16$\,bohr; last two rows of
Table~\ref{tab:pwoo-main}), the AO-only energies agree with the free-space
references to within $0.3$\,mHa, and the cc-pVDZ and cc-pVTZ PWOO
energies lie $2.6$ and $2.7$\,mHa below them, as the STO-3G rows do.  The PWOO
energies are not yet converged in $n_g$ (they change by $-1.6$ and
$-1.9$\,mHa from $n_g=97$ to $129$), and at every grid tested they satisfy the nested-basis bound, lying
below the AO-only energies.

\begin{table*}[t]
\caption{PWOO calculations without padding on H$_8$/SVWN.  The PWOO calculations use $\mathcal B_{\rm PWOO}=\mathcal P\cup\mathcal O$
with $N_{\rm PWOO}=2^{n_q}$.  $E_{\rm AO}$ is the converged $N_{\rm PW}=0$
AO-only reference in the same cell, with the same unit-cell quadrature (Simpson rule, $n_g$ points per
axis) and operator assembly.  Negative
$\Delta_{\rm PWOO-AO}$ satisfies the nested-basis variational bound.  $\Delta_{\rm PWOO-Gauss}$ is $E_{\rm PWOO}$ minus the free-space
Gaussian-basis reference, a comparison with standard molecular
calculations.  $E_{\rm PWOO}$ is the QC-eigensolver SCF result, and $\Delta_{\rm QC\,SCF-cl}$ is its difference from the
classical PWOO SCF for the same calculation.  The last two rows repeat the cc-pVDZ
and cc-pVTZ calculations on the finer grid ($n_g=129$).  Density mixing: linear, $\alpha=0.5$ (STO-3G rows); Anderson,
$\alpha=0.1$, history 6 (cc-pVDZ, cc-pVTZ).  All rows are at $T=0$ with
closed-shell occupations.  The free-space Gaussian-basis references are
$E_{\rm KS}=-4.194417$\,Ha (STO-3G), $-4.284050$\,Ha (cc-pVDZ), and
$-4.315444$\,Ha (cc-pVTZ), computed in free space (no finite cell) with
PySCF~\cite{Sun2018} restricted KS (RKS) calculations with SVWN at the
same $1.4$\,bohr H$_8$ geometry; tighter SCF tolerances and PySCF
integration grids up to level 9 change them by at most $0.04\,\mu$Ha.}
\label{tab:pwoo-main}
\scriptsize
\begin{tabular}{@{}l r r r r r r r r r r r@{}}
\toprule
AO basis & $N_{\rm PW}$ & $N_O$ & $N_{\rm PWOO}$ & $n_q$ & $n_g$ &
$E_{\rm AO}$ (Ha) & $E_{\rm PWOO}$ (Ha) &
$\Delta_{\rm PWOO-AO}$ (mHa) &
$\Delta_{\rm PWOO-Gauss}$ (mHa) &
$\Delta_{\rm QC\,SCF-cl}$ ($\mu$Ha) & Wall \\
\midrule
STO-3G  &  8 &   8 &  16 & 4 &  65 & $-4.189879$ & $-4.196948$ &  $-7.07$ &  $-2.53$ &  $0.00$ & $210$\,s \\
STO-3G  & 24 &   8 &  32 & 5 &  65 & $-4.189879$ & $-4.213935$ & $-24.06$ & $-19.52$ &  $0.00$ & $233$\,s \\
STO-3G  & 56 &   8 &  64 & 6 &  65 & $-4.189879$ & $-4.230440$ & $-40.56$ & $-36.02$ &  $0.00$ & $503$\,s \\
cc-pVDZ & 24 &  40 &  64 & 6 &  65 & $-4.278282$ & $-4.281749$ &  $-3.47$ &  $+2.30$ & $-0.40$ & $383$\,s \\
cc-pVTZ & 16 & 112 & 128 & 7 &  65 & $-4.309203$ & $-4.312849$ &  $-3.65$ &  $+2.59$ & $+0.19$ & $737$\,s \\
\midrule
cc-pVDZ & 24 &  40 &  64 & 6 & 129 & $-4.284105$ & $-4.286666$ &  $-2.56$ &  $-2.62$ & $-0.12$ & $2174$\,s \\
cc-pVTZ & 16 & 112 & 128 & 7 & 129 & $-4.315171$ & $-4.318095$ &  $-2.92$ &  $-2.65$ & $-0.44$ & $3499$\,s \\
\bottomrule
\end{tabular}
\end{table*}

The QC eigensolver does not contribute to these offsets: the QC SCF
energies match the classical PWOO energies on the same grid within
$0.44\,\mu$Ha.

\subsection{Shot-based PWOO readout}
\label{sec:stage5}

We test the shot-based energy estimator on the compact STO-3G
PWOO register ($N_{\rm PWOO}=16$, $n_q=4$) at fixed Hamiltonian.  On the full register, $H_{\rm PWOO}$ is measured
block-wise~\cite{Huggins2021}, as three independent
Hermitian observables in the PWOO basis, the PW--PW block, the
off-diagonal PW--OO plus OO--PW blocks, and the OO--OO block, with the
shots split equally among them.  Each block estimate
has the exact mean and the variance of its observable divided by its number of shots, which is the minimum-variance idealization: it assumes that each
block is measured in its own
eigenbasis.  On the contracted register, energies are estimated
instead from Pauli strings grouped into qubit-wise-commuting (QWC)
sets~\cite{Verteletskyi2020,Crawford2021}, as on hardware.  At fixed Hamiltonian,
the eigensolver using block-wise shot estimates finds the PWOO
eigenvalues to $\mu$Ha accuracy when the residual directions of
the Rayleigh--Ritz step are computed classically from the Hamiltonian
matrix: the maximum eigenvalue error is $4.4\,\mu$Ha at $10^{9}$
shots per batch, against $1295.9\,\mu$Ha without them (on the highest occupied
state).  The shot-SCF runs below do not use these
residual directions.

\begin{figure*}[t]
\centering
\includegraphics[width=0.95\textwidth]{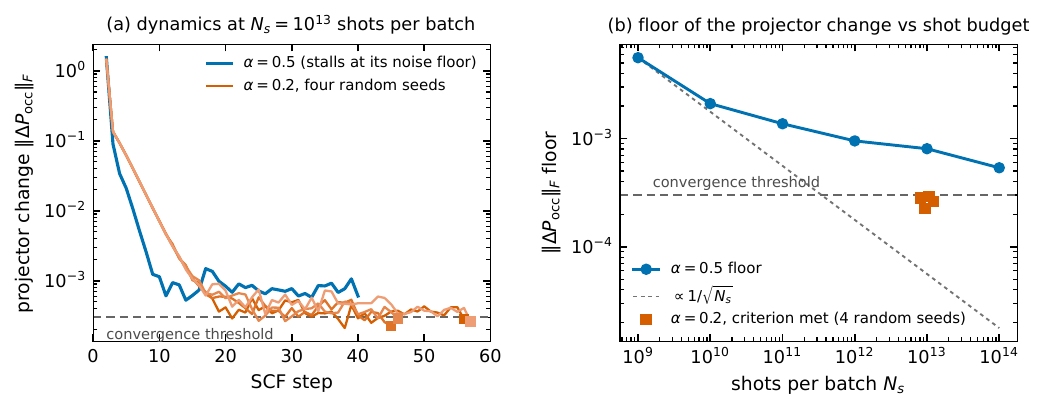}
\caption{Convergence of the shot-based STO-3G PWOO SCF.
(a)~Projector change versus SCF step at $N_s=10^{13}$ shots per batch: with the mixing of the noise-free runs ($\alpha=0.5$) the loop stalls on its noise floor above the convergence threshold
($\|\Delta P_{\rm occ}\|_F=3\times10^{-4}$, dashed), while
a reduced mixing parameter ($\alpha=0.2$) brings $\|\Delta P_{\rm occ}\|_F$ down to the threshold for all four random seeds (squares mark the first steps at which it is below the threshold with $|\delta E|<10^{-5}$\,Ha, i.e., at which the criterion is met;
Table~\ref{tab:pwoo-shot-scf-gate}).  (b)~The $\alpha=0.5$ floor of the projector change (median of $\|\Delta P_{\rm occ}\|_F$ over
the last eight SCF steps) across five decades of shot budget, with the
$1/\sqrt{N_s}$ guide.
The converged $\alpha=0.2$ endpoints at $10^{13}$
shots (squares, horizontally offset for visibility) sit below the
threshold.}
\label{fig:shot-mechanism}
\end{figure*}

Converging the SCF with shot-based eigensolves also requires a smaller
mixing parameter.  With the linear mixing used for
the noise-free STO-3G runs of Table~\ref{tab:pwoo-main} ($\alpha=0.5$)
the shot-SCF stalls.
This is expected when the density update is a fixed-point iteration with
a noisy map: with fixed mixing, noise in each step's output keeps the
iterates fluctuating around the fixed point (a stationary noise floor)
rather than
converging to it, and the mixing parameter $\alpha$ is expected to control
how much of each step's noise enters the next density.  The runs do not
follow this argument in detail ($\alpha=0.1$ does not do better than
$\alpha=0.2$; see below), so the operating point used here was found
empirically.  Two noise sources
contribute: estimator shot noise, which decreases with the shot
budget, and step-to-step variation of the eigensolver output, which we attribute to incomplete convergence of the fixed 128-round schedule with randomized subsets and which does not decrease with the shot budget.  The projector
change therefore falls much more slowly
than $1/\sqrt{N_s}$ ($N_s$ shots per batch; a five-decade sweep from $10^{9}$ to $10^{14}$ shots
per batch lowers the noise floor of the density-matrix change between steps
only $\sim\!7\times$ and
the projector noise floor of Fig.~\ref{fig:shot-mechanism}(b)
$\sim\!10\times$).  In single test runs without randomized subsets (every block updates all
its angles each round), the projector change falls below the threshold
within 17--27 steps, but the SCF converges to energies $0.6$--$2.5$\,mHa above the
noise-free fixed point, so we keep the randomized subsets.  The
40-step $\alpha=0.5$ runs with randomized subsets stall even though their
final energies are within $0.15$--$0.51$\,mHa of the noise-free fixed-point
energy.
At $10^{13}$ shots per batch, reducing the mixing parameter lowers the
fluctuations to just above the threshold, so that the criterion is met on a downward fluctuation: with
$\alpha=0.2$ and $10^{13}$-shot estimator batches (each estimator call
averages eight batches), the PWOO shot-SCF from the initial guess meets the shot-run criterion of Sec.~\ref{sec:method}
($|\delta E|<10^{-5}$\,Ha and $\|\Delta P_{\rm occ}\|_F<3\times10^{-4}$) in $45$--$57$ SCF steps for all four random seeds of the shot sampling that we tested.  At this operating point the projector change fluctuates just above the threshold (median
$3.6$--$3.7\times10^{-4}$ over the last 15 steps), and convergence is
declared at the first step where it falls below $3\times10^{-4}$
while $|\delta E|<10^{-5}$\,Ha.  The final energies lie $+0.05$ to $+0.31$\,mHa above the classical $T=0$ fixed point computed with the same settings
(Table~\ref{tab:pwoo-shot-scf-gate}).  At smaller budgets ($10^{9}$--$10^{11}$ shots) and with $\alpha=0.5$, the
loop instead stops after $18$--$19$ steps under a noise-aware plateau
criterion (stopping when the energy
criterion holds and the mean projector change over the last five steps
is no more than 15\% below its mean over the five before), with
mean energies over the last five steps $0.21$--$0.35$\,mHa above the noise-free
fixed point.
Figure~\ref{fig:shot-mechanism}(a) shows $\|\Delta P_{\rm occ}\|_F$ versus SCF
step, and Fig.~\ref{fig:shot-mechanism}(b) its noise floor versus shot
budget from $10^{9}$ to $10^{14}$ shots per batch.

\begin{table*}[t]
\centering
\caption{Self-consistent PWOO shot-SCF in the STO-3G PWOO basis
($N_{\rm PWOO}=16$, full register $n_q=4$) at
$T=0$ with the mixing parameter reduced to $\alpha=0.2$ and
$10^{13}$-shot estimator batches, for four random seeds of the shot sampling; $\Delta E$ is quoted against the classical $T=0$ fixed point computed
with the settings of the shot runs except linear mixing with $\alpha=0.5$ (the fixed point does not depend on $\alpha$), which agrees with the STO-3G
($N_{\rm PW}=8$) value of Table~\ref{tab:pwoo-main} ($-4.196948$\,Ha;
classical $-4.196948206$\,Ha) to $0.07\,\mu$Ha.  Wall times reflect the
analytic shot emulator (exact means plus sampled noise), whose cost does not depend on the shot count, which is why
$10^{13}$-shot runs can be emulated.}
\label{tab:pwoo-shot-scf-gate}
\footnotesize
\begin{tabular}{@{}l l r r r@{}}
\toprule
Run & Solver / outcome & $E_{\rm KS}$ (Ha) &
$\Delta E$ vs noise-free ($\mu$Ha) & Wall \\
\midrule
Noise-free PWOO fixed point & classical, $T=0$ &
$-4.196948273$ & $0.000$ & $197$\,s \\
Shot-SCF, seed 1234 & criterion met at step 56 &
$-4.196901025$ & $+47.2$ & $824$\,s \\
Shot-SCF, seed 7 & criterion met at step 45 &
$-4.196808047$ & $+140.2$ & $676$\,s \\
Shot-SCF, seed 8 & criterion met at step 46 &
$-4.196640804$ & $+307.5$ & $669$\,s \\
Shot-SCF, seed 9 & criterion met at step 57 &
$-4.196878734$ & $+69.5$ & $825$\,s \\
\bottomrule
\end{tabular}
\end{table*}

These are simulated shot draws; the real-device readout experiment
is reported in Sec.~\ref{sec:hardware}.  The emulated readout tests show that block-wise shot estimates, with residual directions computed from the
Hamiltonian matrix, reproduce the fixed-Hamiltonian eigenvalues to $4.4\,\mu$Ha
at $n_q=4$, and that with $\alpha=0.2$ the shot-based SCF meets the shot-run criterion on the same register.  Among the runs with randomized subsets, only $\alpha=0.2$ at $10^{13}$ shots per batch met the criterion within the step limits: at
$10^{11}$--$10^{12}$ shots, $\alpha=0.2$ left the projector change above
the threshold, near the $\alpha=0.5$ floor at the same shot budget; at $10^{13}$ shots,
$\alpha=0.1$ had still not reached the threshold after 120 steps.

\subsection{Contracted PWOO on the three-qubit register used for pure PW}
\label{sec:pwoo-contraction}

The dual-basis contraction of Sec.~\ref{sec:contraction} applies to
the PWOO basis as it does to pure PW (the settings that differ are listed in Appendix~\ref{app:details}).  After a
full-register bootstrap solve, the $K$ lowest QC eigenstates form the
contracted basis $U$; subsequent SCF steps solve
$\tilde H = U^{\dagger} H_{\rm KS}[\rho]\,U$ on the fixed
$n_q'=\lceil \log_2 K\rceil$ register; the lifted orbitals form the PWOO density; and the residual norm $\eta_m$ triggers subspace updates at the
same threshold, $2.5\times10^{-3}$\,Ha.
Table~\ref{tab:pwoo-contracted} reports the resulting runs.  On the
cc-pVTZ PWOO row the contracted solves use $n_q'=3$ qubits instead of the
full register's $n_q=7$: the converged energies agree with the uncontracted
QC row to $4.7$--$35.1\,\mu$Ha, and each contracted solve takes $0.3$\,s against $3$--$21$\,s per
full-register solve; $29$ of the $39$ SCF steps of the block-parallel state-vector row use only contracted solves, and the other $10$ are the bootstrap
step and the $9$ steps with a subspace update.  Serial (Gauss--Seidel) and block-parallel Rotosolve both give converged SCF runs at the two register sizes tested;
block-parallel Rotosolve is $1.2\times$ slower at $n_q=4$ (STO-3G) and
$1.7\times$ faster at $n_q=7$ (cc-pVTZ) in total SCF wall time; two sizes
do not establish a scaling trend.

\begin{table*}[t]
\caption{Contracted PWOO SCF on H$_8$/SVWN at $K\!=\!8$
($n_q'\!=\!3$), residual threshold $2.5\times10^{-3}$\,Ha.  $\Delta E$ is quoted against the uncontracted calculation with the same
settings (STO-3G: classical fixed point
$-4.196948$\,Ha; cc-pVTZ: the QC row of
Table~\ref{tab:pwoo-main}, $-4.312849$\,Ha, whose uncontracted
run took 33 steps / 737\,s).  SCF mode: classical = classical eigensolver; SV = QC eigensolver with
state-vector expectation values; shots = QC eigensolver with simulated
measurement.  Steps and $n_{\rm upd}$ as in
Table~\ref{tab:main-results}.  The shot row runs every
contracted-register eigensolve with the $10^{13}$-shot estimator grouped
into QWC Pauli sets and linear mixing with $\alpha=0.2$ (as in
Table~\ref{tab:pwoo-shot-scf-gate}) in place of the Anderson mixing of the
noise-free cc-pVTZ rows; bootstrap
and subspace-update solves on the full register are state-vector, so shots
enter only on the contracted register.
Contraction preserves the energy relations of Table~\ref{tab:pwoo-main}: every
STO-3G row lies below the free-space STO-3G Gaussian-basis reference
($-4.194417$\,Ha), and every cc-pVTZ row lies $2.6$\,mHa above the free-space cc-pVTZ
Gaussian-basis reference ($-4.315444$\,Ha), the $n_g=65$ grid offset of
Table~\ref{tab:pwoo-main}.}
\label{tab:pwoo-contracted}
\footnotesize
\begin{tabular}{@{}l l l c c c r r@{}}
\toprule
AO basis ($N_{\rm PWOO}\!\to\!K$) & Rotosolve & SCF mode & steps &
$n_{\rm upd}$ & Wall & $E_{\rm KS}$ (Ha) & $\Delta E$ ($\mu$Ha) \\
\midrule
STO-3G ($16\!\to\!8$) & --- & classical & 23 & 5 & ---
& $-4.196943$ & $+5.1$ \\
STO-3G ($16\!\to\!8$) & serial (Gauss--Seidel) & SV & 23 & 5 & $203$\,s
& $-4.196943$ & $+5.1$ \\
STO-3G ($16\!\to\!8$) & block-parallel & SV & 23 & 5 & $252$\,s
& $-4.196943$ & $+5.1$ \\
cc-pVTZ ($128\!\to\!8$) & serial (Gauss--Seidel) & SV & 35 & 9 & $1289$\,s
& $-4.312843$ & $+5.7$ \\
cc-pVTZ ($128\!\to\!8$) & block-parallel & SV & 39 & 9 & $741$\,s
& $-4.312844$ & $+4.7$ \\
cc-pVTZ ($128\!\to\!8$) & block-parallel & shots ($10^{13}$, contracted register)
& 46 & 11 & $2248$\,s & $-4.312814$ & $+35.1$ \\
\bottomrule
\end{tabular}
\end{table*}

The last row of Table~\ref{tab:pwoo-contracted} is an SCF in the cc-pVTZ PWOO basis in which all contracted-register
eigensolves use energy estimates that each average eight $10^{13}$-shot
batches and the full-register
solves use state vectors, on the same three-qubit register used by the $n_q=15$ pure-PW calculation.  It meets the shot-run criterion (final projector
change $2.1\times10^{-5}$ against the $3\times10^{-4}$ threshold).  Its
contracted solves use no randomized subsets (the source of the
eigensolver-output noise in Sec.~\ref{sec:stage5}), and its contracted
$\tilde H$ acts on 3 qubits: the standard deviation of
each energy estimate from $N_s$ shots is of order $\|h\|_1/\sqrt{N_s}$ (at
most $\sqrt{N_{\rm set}\sum_g\|h_g\|_1^2/N_s}$ for the equal split over the $N_{\rm set}$ QWC
sets used here), with $\|h\|_1=0.68$\,Ha for this
contracted PWOO Hamiltonian throughout the SCF ($0.59$--$0.75$\,Ha for the
pure-PW contractions).  End to end, the contracted-register shot run makes at least $1.5\times10^{5}$ estimator calls ($4{,}361$
Rotosolve rounds over $34$ contracted SCF steps), each averaging eight
$10^{13}$-shot batches, or more than $10^{19}$ simulated shots in
total; the full-register solves of the bootstrap step and of the $11$
steps with a subspace update ran in state-vector mode.  These counts give the emulated noise level at which this run converged;
we did not determine the largest tolerable noise, and they are not a
hardware shot budget.  The per-evaluation bound of Sec.~\ref{sec:stage2}
($\sim\!10^5$ shots for one energy to chemical accuracy) does not set this
budget, which reflects the precision needed by the angle optimization and
the SCF.  In this run the $n_q=7$ full-register solves are
state-vector; shot-based full-register eigensolves were run only on the
$n_q=4$ STO-3G register (Sec.~\ref{sec:stage5}).

\subsection{Hardware readout on a trapped-ion quantum processor}
\label{sec:hardware}

\begin{figure*}[t]
\centering
\includegraphics[width=0.95\textwidth]{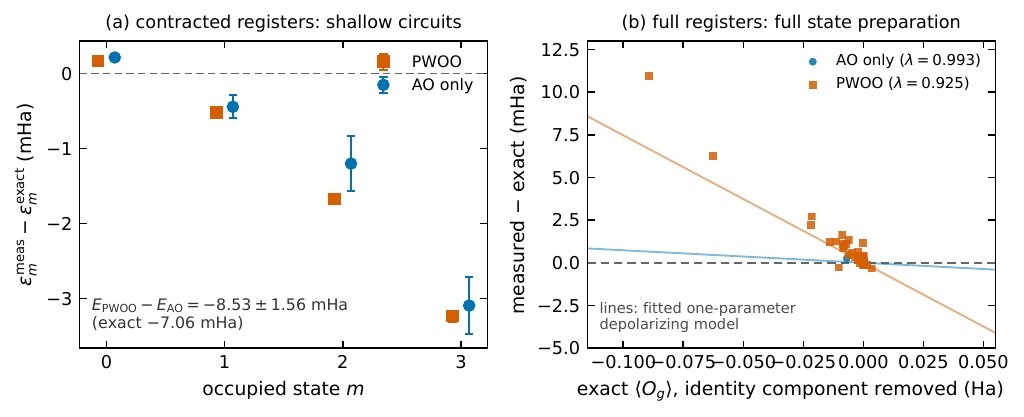}
\caption{Trapped-ion readout of the converged PWOO KS states (IonQ
Forte-1, debiased, two repetitions).  (a)~Contracted three-qubit
registers: per-state hardware bias for the PWOO and AO-only states
(offset horizontally from integer $m$ for visibility; error bars:
standard deviation over the two repetitions).  (b)~Full registers ($11$ vs $4$ two-qubit gates): per-set deviation of the measured $\langle O_g\rangle$ from the exact values.  Lines: fitted one-parameter depolarizing model ($\lambda=0.993$ for
AO-only, $0.925$ for PWOO).}
\label{fig:hardware}
\end{figure*}

In a real-device experiment on IonQ Forte-1 (36
qubits~\cite{IonQForte}; see Ref.~\cite{Chen2024Forte} for a benchmark of
the Forte system), no optimization is performed on the device: the converged
KS eigenstates of the STO-3G PWOO calculation ($N_{\rm PW}=8$, from an
earlier run of that row; Appendix~\ref{app:details}) and of its AO-only
reference are
prepared as circuits, and their band energies $\sum_mf_m\epsilon_m$ (on the full registers, only the lowest eigenvalue) are measured by Pauli
readout grouped into QWC sets, two full
repetitions each, with IonQ's
debiasing error mitigation (symmetrized variants of each circuit are
run and their results aggregated)~\cite{Maksymov2023}.
Every job is archived with its Forte-1 job identifier.  We measure on the contracted three-qubit registers and on the full
registers ($n_q=4$ for PWOO, $n_q=3$ for AO-only; Fig.~\ref{fig:hardware}).

The circuits are prepared from the converged amplitudes with Qiskit's
state-preparation routine, a uniformly controlled rotation construction of
the same type as the M\"ott\"onen circuit, and transpiled for the device.
On the contracted three-qubit registers each measured state is, up to a
phase, a computational basis state $|m\rangle$, because $U$ is built from
the measured orbitals themselves; $\tilde H$, assembled at the converged
density, is nearly diagonal.  At other densities $\tilde H$ is not
diagonal and its eigenstates are general three-qubit states, and
$\eta_m$ tests whether ${\rm span}(U)$ still represents them.  The
contracted circuits have at most $4$ two-qubit gates (none for the
lowest PWOO orbital) because Qiskit's general state-preparation routine
is used; a computational basis state needs none.  This test therefore
probes the measurement of $\tilde H$, not the preparation of general
states.  No noise-model calibration was applied to the contracted-register data.  The measured energies are $E=E_{\rm KS}^{\rm exact}+\sum_m f_m(\epsilon_m^{\rm meas}-\epsilon_m^{\rm exact})$:
the band-energy part $\sum_m f_m\epsilon_m$ is measured on hardware and
the double-counting terms are evaluated classically at the converged
density.  \emph{Exact} denotes noise-free classical values for the same states
and Hamiltonian: $\epsilon_m^{\rm exact}=\langle\psi_m|H_{\rm KS}[\rho^*]|\psi_m\rangle$
with $\rho^*$ the converged density ($\tilde H$ in place of $H_{\rm KS}$ on
the contracted registers), and $E_{\rm KS}^{\rm exact}$ is the
SCF energy of the source calculation.  The measured band-energy difference is $-62.3$\,mHa against
$-60.9$\,mHa exact, and the total-energy difference,
$E_{\rm PWOO}-E_{\rm AO} = -8.53 \pm 1.56$\,mHa (standard deviation over
the two repetitions), compares with the exact $-7.06$\,mHa for the
measured states, within one standard deviation
[Fig.~\ref{fig:hardware}(a)].  The per-state hardware biases grow in magnitude with the orbital index,
from $+0.2$\,mHa for the lowest orbital to $-3.2$ and $-3.1$\,mHa for the
highest occupied one (PWOO and AO-only, respectively), and are nearly equal
between the PWOO and AO-only states, so they largely cancel in the PWOO--AO difference; we therefore report
the difference.

On the full registers the state preparations are complete
amplitude-encoding circuits ($11$ two-qubit gates for the
$16$-dimensional PWOO basis, $4$ for the $8$-dimensional AO
basis),
and without calibration the lowest-eigenvalue difference (only the
lowest orbital is measured on the full registers) is
$+52.4 \pm 2.7$\,mHa against $-7.54$\,mHa exact, because the PWOO circuits carry $+66$\,mHa of hardware bias against
$+6$\,mHa for the AO-only circuits.
The statistical precision ($2.7$\,mHa over $5.4$--$5.5\times10^{5}$
shots per system and repetition, allocated across the QWC sets in proportion to each set's standard deviation, with at least $500$ shots per set) is much smaller than this 60\,mHa difference in hardware bias.
We apply a one-parameter depolarizing
calibration~\cite{Vovrosh2021,Urbanek2021}: a single rescaling factor
$\lambda$ per system and repetition, shared by all its QWC-set circuits,
is fit by weighted least squares to the measured expectation values
$\langle O_g\rangle$ of the QWC sets $g$ (each $O_g$ is the sum of the
Pauli terms measured together, without its identity component), so that
the remaining degrees of freedom test the model
[Fig.~\ref{fig:hardware}(b)].  This calibration brings the AO-only readout into agreement with the exact value
($\lambda=0.993$, calibrated energy within $0.2$\,mHa of exact) but
not the PWOO readout: the fit gives $\lambda=0.925$ with $\chi^2/{\rm dof}=3.5$ and
$5.6$ in the two repetitions and leaves a $+10$ and $+14$\,mHa
bias beyond the depolarizing model in the lowest eigenvalue, so the calibrated difference, $+4.2\pm2.2$\,mHa (propagated uncertainty of the
calibration fit), has the wrong sign.

More elaborate mitigation (richer noise models,
zero-noise extrapolation~\cite{Temme2017,Li2017}) or shorter state preparations may recover
the full-register readout; in the present data, only the contracted-register measurements
reproduce the exact difference.

\subsection{Discussion}
\label{sec:discussion}
\label{sec:senjean-comparison}

The closest prior simulated full KS-loop quantum-computing demonstration is
Senjean et al.~\cite{Senjean2022}, also on H$_8$/SVWN.  Their
calculation uses an STO-3G AO basis, a hardware-efficient
ansatz~\cite{Kandala2017}, and subspace-search (ensemble) VQE multi-orbital
optimization~\cite{Nakanishi2019}, and includes SCF runs with sampling
noise.  They report H$_8$ energies for spacings of 1.0--3.0\,\AA{}, so no direct
numerical comparison is possible at our $1.4$\,bohr ($0.74$\,\AA) spacing,
where the STO-3G PWOO energies lie below the free-space STO-3G energy
because PWOO adds plane waves to the STO-3G basis.  In contrast to that
work, the $n_q=15$ pure-PW calculation shows, in state-vector emulation, an SCF in a $32\,768$-function
all-electron plane-wave basis, a basis that can be enlarged
systematically to any energy tolerance; the QC eigensolver solves all
its steps except the bootstrap steps and subspace updates on a
three-qubit contracted register.  PWOO falls below the STO-3G reference
energy with 16 functions, whereas pure PW needs more than 4096, and the nested-basis variational bound checks the assembly of its
PW--AO operator blocks.

Plane waves are standard in DFT and label the computational basis states
of first-quantized quantum algorithms~\cite{Babbush2019,Su2021}, and
Senjean et al.\ already encode an $N$-dimensional KS matrix on
$\log_2N$ qubits and suggest plane waves for it~\cite{Senjean2022}.  To our knowledge, what is new here is an all-electron plane-wave and PWOO basis in a self-consistent
QC KS loop, on a register whose width stays fixed as the full basis
grows.

The SCF loop runs in state-vector or simulated-shot mode, and the
hardware experiment measures the converged states on the contracted and
full registers.  In these emulations the orbitals are rebuilt classically
from the circuit angles, so the classical side holds all amplitudes of
each orbital and uses them for the deflation overlaps, MGSO, and the
density.  On hardware, the overlaps would have to be estimated on the
device, for example by swap or Hadamard tests, and rebuilding the
density from the angles scales with the basis size; we therefore make no
claim of a quantum speed-up.  The implementation is spin-restricted SVWN.  The PWOO rows with the larger AO bases satisfy the nested-basis
variational bound; at $n_g=65$ they lie 2.3--2.6\,mHa above their
free-space Gaussian-basis references, an offset of the unit-cell grid, and
at $n_g=129$ they lie 2.6 and 2.7\,mHa below them.  The SCF with shot-based eigensolves converges in the compact STO-3G
PWOO basis on its full register and, in
contracted form, in the cc-pVTZ PWOO basis (shots on the contracted
register only; shot-based full-register solves beyond the $n_q=4$
register of Table~\ref{tab:pwoo-shot-scf-gate} were not attempted and
remain to be tested);
the noise-aware stopping rule used at the
smaller shot budgets stops on a plateau of the projector change rather than a fixed
tolerance and has not been validated as a convergence criterion.  Spin polarization, generalized-gradient approximations, hybrid
functionals, and broader PWOO basis sweeps are planned.

\section{Conclusions}
\label{sec:conclusions}

We demonstrated a self-consistent KS-DFT method in which the inner
eigenvalue problems are solved by a quantum-circuit eigensolver that
accepts the KS matrix in any orthonormal basis.  The pure-PW calculation
at $n_q=15$ ($K=8$) uses a three-qubit contracted register and agrees with the same-basis classical PW reference to $0.015$\,mHa.
PWOO reduces the basis size needed for a given energy, and its
construction implies a nested-basis variational bound, which we use as a
consistency check.  Because PWOO changes only how the KS matrix is
assembled, the eigensolver runs on it unchanged, and because it needs
far fewer functions for a given energy it reduces the width of the
full-register solves.  The STO-3G, cc-pVDZ, and cc-pVTZ PWOO calculations satisfy that bound on
full registers of $n_q=4$--7 qubits, and the PWOO SCF
calculations with the QC eigensolver match their classical counterparts
to within $0.44\,\mu$Ha.  The noise-free block-parallel solves include classical residual
directions in their Rayleigh--Ritz step; the shot-based SCF eigensolves
omit them, and the Gauss--Seidel runs (pure PW at $n_q\le12$, the
contracted solves at $n_q=15$, and the serial rows of
Table~\ref{tab:pwoo-contracted}) have no Rayleigh--Ritz step.  For the STO-3G PWOO basis ($n_q=4$),
the SCF with shot-based eigensolves at $10^{13}$ shots per batch and
$\alpha=0.2$ meets the shot-run convergence criterion, with a projector
change that fluctuates just above the threshold until it first falls
below it, and with energies $0.05$--$0.31$\,mHa above the noise-free fixed
point; at $\alpha=0.5$ with randomized subsets the projector change stays above the threshold.  The same contraction applies to PWOO, so pure PW at
$N_{\rm PW}=32\,768$ and PWOO with the STO-3G and cc-pVTZ AO bases solve all SCF
iterations except the bootstrap and subspace updates on the same
three-qubit register, including a run whose contracted-register solves are all shot-based.  The STO-3G PWOO energies lie below the free-space STO-3G Gaussian-basis
reference, and
on the finer unit-cell grid ($n_g=129$) the cc-pVDZ and cc-pVTZ PWOO
energies lie $2.6$ and $2.7$\,mHa below their free-space references; these offsets
are not yet converged in $n_g$, whereas the nested-basis bound is satisfied at
every grid tested.  On IonQ Forte-1, readout of the converged STO-3G states on the
contracted three-qubit registers reproduces the PWOO--AO energy
difference within one standard deviation of two repetitions; full-register
readout does not, even after depolarizing calibration.

\section*{Data Availability}
The Forte-1 measurement records (per-job identifiers, shot counts,
and histograms) are archived with the analysis script.
The code for the PWOO basis and its integrals, the basis contraction, the
M\"ott\"onen ansatz, the Rotosolve eigensolver with penalty deflation and
MGSO, the shot estimators, and the SCF loop are implemented in a self-contained
Python package
(NumPy/SciPy, with qiskit-ionq for circuit submission), which comes with
the driver scripts and, where available, the result JSON files
underlying the numerical results.
Unless explicitly marked total, archived JSONs report the
electronic-only $F$ and $E$ (without $V_{\rm NN}$).
Total $E_{\rm KS}$ values quoted in the paper
add the H$_{8}$ nuclear-repulsion energy
$V_{\rm NN}\!=\!+9.8163265$\,Ha at $1.4$\,bohr spacing. The archived $n_q\!=\!9$ $K$-sweep files are classical-reference runs
(classical bootstrap and contracted solver) at the same residual threshold; at $K\!=\!8$ the corresponding QC-eigensolver run agrees with
them to $0.02$\,mHa.  The pure-PW calculations of
Table~\ref{tab:main-results} and the PWOO calculations of
Table~\ref{tab:pwoo-main} are archived as JSON/NPZ files, together with their classical counterparts.
The settings needed to reproduce each calculation are listed in
Appendix~\ref{app:details}.  All code and result files are available from the author on
reasonable request; a public repository release is planned to
accompany the journal version.

\appendix

\section{Computational details}
\label{app:details}

This appendix lists the settings needed to reproduce the calculations.
Energies are in Ha and lengths in bohr throughout.

\subsection{System and Coulomb kernel}

The eight H nuclei ($Z=1$, bare point charges) sit at
$x_i=(i-\tfrac72)\times1.4$, $y_i=z_i=0$ ($i=0,\ldots,7$) in the cube
$[-L/2,L/2]^3$, $L=20$, for the pure-PW calculations; the PWOO
calculations use the same positions with the chain along $z$.  No
nucleus lies on a grid point.  The nuclear repulsion is
$V_{\rm NN}=9.8163265$.  The functional is spin-restricted SVWN: Slater
exchange and VWN5 correlation~\cite{Vosko1980} (libxc LDA\_X and
LDA\_C\_VWN~\cite{Lehtola2018}, evaluated through
PySCF~\cite{Sun2018,Sun2020}; PySCF's
\texttt{svwn} maps to the same pair).  The spherically truncated Coulomb kernel~\cite{Jarvis1997,Rozzi2006} used
for the electron--nuclear and Hartree terms is
\begin{equation}
v(\mathbf G)=\frac{4\pi}{G^2}\bigl(1-\cos GR_c\bigr)\quad(G\neq0),\qquad
v(\mathbf 0)=2\pi R_c^2,
\label{eq:mt-kernel}
\end{equation}
with $R_c=L/2=10$.

\subsection{Pure-PW calculations (Table~\ref{tab:main-results})}

\emph{Basis and grid.}  $\mathbf G=(2\pi/L)\mathbf m$ with
$m_x,m_y,m_z\in\{-n/2,\ldots,n/2-1\}$ (the full cube, $N_{\rm PW}=n^3$,
$n=4,8,16,32$), kinetic energy $\frac12|\mathbf G|^2$, and
$H_{\mathbf G\mathbf G'}=\frac12|\mathbf G|^2\delta_{\mathbf G\mathbf G'}
+\tilde V_{\rm KS}(\mathbf G-\mathbf G')$, where $\tilde V_{\rm KS}$ is the
discrete Fourier transform of $V_{\rm KS}(\mathbf r)$ on the $n^3$
real-space grid $r_j=jL/n-L/2$ and $\mathbf G-\mathbf G'$ is taken modulo
$n$.  The density, Hartree, exchange--correlation, and external
potentials are all evaluated on this $n^3$ grid; $V_{\rm en}(\mathbf r_j)$
is the real part of the inverse discrete transform, over the same cube of
$\mathbf G$, of $-\Omega^{-1}v(\mathbf G)\sum_Ie^{-i\mathbf G\cdot\mathbf R_I}$
multiplied by $(-1)^{m_x+m_y+m_z}$ for the grid origin at $-L/2$, with no
special treatment of the $m=-n/2$ components.  Basis state $|q\rangle$
is the $q$-th plane wave in order of increasing $|\mathbf G|^2$, with ties
broken by $m_x$, then $m_y$, then $m_z$; the root of the M\"ott\"onen tree
is the most significant bit of $q$.

\emph{SCF.}  The initial density is a sum of normalized Gaussians
$e^{-r^2/2\sigma^2}/(2\pi\sigma^2)^{3/2}$ with $\sigma=1$ on the nuclei
(minimum-image distances), rescaled to eight electrons.  The mixed
quantity is $\rho(\mathbf r)$ on the grid; the first step is linear and
later steps use Anderson (type-II) mixing, $\rho_{t+1}=\rho_t+\alpha
r_t-\sum_i\gamma_i[(\rho_t-\rho_i)+\alpha(r_t-r_i)]$ with
$r=\rho_{\rm out}-\rho_{\rm in}$ and $\gamma$ from a least-squares fit
(Euclidean norm on the grid) over the stored pairs (at most the history length, including the current pair); the history is kept across subspace
updates.  Fermi--Dirac
occupations are $f=2/(1+e^{(\epsilon-\epsilon_F)/k_BT})$ over the $K=8$ solved
states, with the Fermi level $\epsilon_F$ from a
bracketed root search for $\sum_mf_m=8$; the free energy is $F=E_{\rm KS}+2k_BT\sum_m[p_m\ln p_m+(1-p_m)\ln(1-p_m)]$,
with $p=f/2$.  The energy is
$E_{\rm KS}=\sum_mf_m\epsilon_m-\int\rho_{\rm out}(v_{\rm H}+v_{\rm xc})[\rho_{\rm in}]\,d\mathbf r
+E_{\rm H}[\rho_{\rm out}]+E_{\rm xc}[\rho_{\rm out}]+V_{\rm NN}$.
Convergence requires $|F_t-F_{t-1}|<10^{-5}$ and
$\max_{\mathbf r}|\rho^t_{\rm out}-\rho^{t-1}_{\rm out}|<10^{-4}$\,bohr$^{-3}$.
Per register size ($n_q=6,9,12,15$): mixing $\alpha$/history $0.1/6$,
$0.3/2$, $0.3/2$, $0.3/2$; bootstrap steps 2, 3, 3, 5; occupations pinned
$[2,2,2,1,1,0,0,0]$ at $T=0$ for $n_q=6$ and Fermi--Dirac with
$k_BT=3$\,mHa otherwise.  Table~\ref{tab:main-results} reports $E_{\rm KS}$.

\emph{Contraction.}  After the bootstrap, $U$ is the orthonormal factor of the QR decomposition of the eight lowest full-basis eigenvectors, and $\tilde H=U^\dagger H U$ is
solved for all $K=8$ states.  $\eta_m$ is evaluated with the undeflated
$H$ for the four lowest states (at $n_q=6$ this excludes the fifth,
singly occupied state); when $\max_m\eta_m>2.5\times10^{-3}$, the
full-basis problem is solved in the same step, warm-started from the
lifted states, $U$ is replaced by the orthonormal QR factor of the new eigenvectors,
and $\tilde H$ is solved again at the same density.  ``Steps'' counts
bootstrap and contracted steps; $n_{\rm upd}$ counts these updates.
The classical references use Lanczos (\texttt{scipy.sparse.linalg.eigsh}~\cite{Virtanen2020,Lehoucq1998},
eight lowest states, \texttt{tol=0}) on the FFT matrix--vector product,
with the same SCF settings; the classical $K=8$ runs use it for the
bootstrap and updates and dense diagonalization of $\tilde H$.  The
Fig.~\ref{fig:basis-conv} points at $n_q=18$ and $21$ ($E_{\rm KS}=-4.309446$
and $-4.317414$) are uncontracted classical Lanczos SCF runs at $T=0$
(integer occupations) with $|\delta F|<10^{-6}$ and
$\max|\Delta\rho|<10^{-5}$; the same series reproduces the
Table~\ref{tab:main-results} references (the circles at $n_q\le15$ in
Fig.~\ref{fig:basis-conv}) at $n_q=12$ and 15 to within $1\,\mu$Ha ($17\,\mu$Ha at $n_q=9$, from the different occupations).

\emph{Pauli L1 norms.}  The full and contracted L1 norms are both evaluated at the
converged density of classical $K=8$ runs (script and data archived) with Fermi--Dirac occupations
at $k_BT=3$\,mHa for all three sizes; for $\tilde H$ the columns of $U$ are the eight
lowest eigenvectors in ascending order, so that $\tilde H$ is diagonal.  Both include
the identity coefficient and all coefficients above $10^{-10}$, and the
shot bound is $(\|h\|_1/\delta)^2$ with $\delta=1.6$\,mHa.

\subsection{PWOO calculations (Tables~\ref{tab:pwoo-main}--\ref{tab:pwoo-contracted})}

\emph{Basis.}  The candidate plane waves are $\mathbf m\in\{-5,\ldots,4\}^3$,
ordered by $|\mathbf G|^2$ with ties broken by $m_x$, $m_y$, $m_z$; the
first $N_{\rm PW}$ are kept (these sets do not keep $\pm\mathbf G$ pairs
together).  The AO bases are the PySCF STO-3G, cc-pVDZ, and cc-pVTZ sets
with spherical harmonics ($N_{\rm AO}=8$, 40, 112).  All overlaps and the
kinetic blocks are analytic integrals over the cell, including the PW--AO
overlaps $\Omega^{-1/2}\int_{\rm cell}q_a\,e^{-i\mathbf G\cdot\mathbf r}\,d\mathbf r$;
the AO--AO kinetic block is $\frac12\int_{\rm cell}\nabla q_a\cdot\nabla
q_b\,d\mathbf r$, and the PW--AO kinetic block is $\frac12|\mathbf G|^2$
times the PW--AO overlap.  For the cell-restricted AOs, the latter omits
the surface term $\frac12\oint_{\partial\,{\rm cell}}q_a\,\partial_n
P_{\mathbf G}^*\,dS$, which the form
$\frac12\int_{\rm cell}\nabla P_{\mathbf G}^*\cdot\nabla q_a\,d\mathbf r$
includes; it is largest for the most diffuse cc-pVDZ and cc-pVTZ
functions.  Including it lowers the classical PWOO energies of
Table~\ref{tab:pwoo-main} by $0.02$--$0.14$\,mHa and leaves the AO-only
references, and hence the nested-basis bound, unchanged; the comparisons
reported here (QC versus classical, contracted versus uncontracted, shots
versus the noise-free fixed point) use the same matrix on both sides.  Canonical orthonormalization, first of the AOs and then of the residual
overlap matrix $1-A^\dagger A$ (with $A$ the PW--AO overlaps), keeps
eigenvalues above $10^{-10}\max(\lambda_{\max},1)$; no function is dropped
in any calculation.  Basis index $j$, with the plane waves first and then
the OO functions in order of decreasing eigenvalue of the residual
overlap matrix, is register state $|j\rangle$.  For the AO-only basis
($N_{\rm PW}=0$) the residual overlap matrix is the identity and the
register order is that of the canonical AO vectors in order of decreasing
overlap eigenvalue; basis-vector phases are those returned by the
Hermitian eigensolver of NumPy (LAPACK).

\emph{Operators.}  The kinetic blocks are given in Sec.~\ref{sec:pwoo-main}.
The electron--nuclear PW--PW block uses Eq.~(\ref{eq:mt-kernel}) with the
nuclear structure factor, the AO--AO block the free-space Gaussian
integrals, and the PW--AO block a unit-cell Simpson quadrature of the bare
$-Z/|\mathbf r-\mathbf R|$.  For the Hartree matrix, the Fourier coefficients
$\sigma_{ij}(\mathbf q)=\int_{\rm cell}\chi_i^*\chi_je^{-i\mathbf q\cdot\mathbf r}d\mathbf r$
of all products of raw PW and AO functions are computed analytically for
$\mathbf q=(2\pi/L)\mathbf m$, $\mathbf m\in\{-n/2,\ldots,n/2-1\}^3$, and
$V_{ij}=\Omega^{-1}\sum_{\mathbf q}v(\mathbf q)\rho(\mathbf q)\sigma_{ij}(-\mathbf q)$.
The raw matrix is extrapolated between two cutoffs,
$M=(n_2^2M_2-n_1^2M_1)/(n_2^2-n_1^2)$, with $(n_1,n_2)=(8,12)$ for
$N_{\rm PW}\le24$ and $(10,14)$ for $N_{\rm PW}=56$, and the AO--AO block
generated by the AO--AO density is replaced by the free-space Coulomb
matrix; $E_{\rm H}=\frac12{\rm Tr}(DV_{\rm H})$.  For exchange--correlation, the
density is evaluated on the unit-cell grid, clamped at zero, and rescaled
to eight electrons; $E_{\rm xc}$ is summed on that grid, and the PW--PW block is
$\Omega^{-1}v_{\rm xc}(\mathbf G-\mathbf G')$ with
$v_{\rm xc}(\mathbf q)=\sum_{\mathbf r}w_{\mathbf r}v_{\rm xc}(\mathbf r)e^{-i\mathbf q\cdot\mathbf r}$
summed over the same grid ($\mathbf q$ on the cube with $n=8$, or 10 for
$N_{\rm PW}=56$).  The blocks involving AOs use atom-centered quadrature
with the density evaluated analytically from $D$, clamped at zero, and
multiplied by the unit-cell rescaling factor (so that it integrates to
eight electrons on the unit-cell grid, not on the atom-centred grid): 129 uniform radial points on $[0,9.5]$ with Simpson weights,
the 110-point Lebedev angular grid~\cite{Lebedev1999}, Becke partition weights~\cite{Becke1988} (three
iterations, no atomic-size adjustment), and points outside the
cell discarded without renormalizing the weights.  The unit-cell grid has $n_g$ points per axis
on $[-L/2,L/2]$, endpoints included (spacing $L/(n_g-1)$), with Simpson
weights.

\emph{SCF.}  The mixed quantity is the density matrix
$D=\sum_mf_m\,c_mc_m^\dagger$ in the PWOO basis ($c_m$ the PWOO coefficient vector of orbital $m$), with
$E_{\rm KS}={\rm Tr}(Dh_{\rm core})+E_{\rm H}[D]+E_{\rm xc}[D]+V_{\rm NN}$ ($h_{\rm core}$ the
kinetic plus electron--nuclear matrix)
evaluated at the mixed density of each step (the input density of the
next step).  Mixing is linear with $\alpha=0.5$ for
the STO-3G PWOO rows and the noise-free STO-3G contracted rows, and Anderson (type II, as above) with $\alpha=0.1$ and
history 6 for the cc-pVDZ and cc-pVTZ rows, the AO-only references, and
the noise-free cc-pVTZ contracted rows; the contracted shot row uses linear
mixing with $\alpha=0.2$.  In the Anderson fit for $D$, $\gamma$ minimizes
the Frobenius norm, $\|R\|_F^2={\rm Tr}(R^\dagger R)$; because the residual
differences are Hermitian, the fitted $\gamma$ is real, and the mixed $D$ is
re-Hermitized.  Occupations are $[2,2,2,2,0,0,0,0]$ at $T=0$
for $K=8$ solved states.  The initial density matrix is built from the four lowest eigenvectors of
$h_{\rm core}$, obtained by classical diagonalization, and the first
eigensolve is warm-started from the eight lowest; later eigensolves are warm-started from the
previous step's orbitals.  The maximum number of SCF steps is 150 for the noise-free PWOO runs of
Tables~\ref{tab:pwoo-main} and~\ref{tab:pwoo-contracted} (60 for the
noise-free fixed point of Table~\ref{tab:pwoo-shot-scf-gate}) and 80--120
for the shot runs: 80 for $\alpha=0.2$, 120 for $\alpha=0.1$, 100 for the
contracted shot row, and 40 for all $\alpha=0.5$ shot runs.  The shot runs
use seed 1234, plus seeds 7, 8, and 9 for the $\alpha=0.2$, $10^{13}$-shot
runs of Table~\ref{tab:pwoo-shot-scf-gate}.  At
$n_g=97$ the classical PWOO energies are $-4.285075$ (cc-pVDZ) and
$-4.316235$ (cc-pVTZ), and the AO-only energies $-4.194346$ (STO-3G),
$-4.283200$, and $-4.314040$.  The
classical PWOO eigensolver is dense diagonalization of $H_{\rm PWOO}$.  In
the contracted PWOO runs (Table~\ref{tab:pwoo-contracted}) there is one
bootstrap step, $U$ is the orthonormal QR factor of the eight QC orbitals, $\eta_m$ is
evaluated for all eight states, the contracted solves are warm-started
from the previous orbitals projected onto $U$, and the noise-free rows
use the three-step PWOO criterion.

\emph{Gaussian-basis references.}  PySCF RKS with \texttt{xc="svwn"}, the
default integration grid (level 3), and \texttt{conv\_tol}$=10^{-9}$, at
the same geometry in free space.

\subsection{QC eigensolver}

For $n_q$ qubits the ansatz has $2^{n_q}-1$ amplitude angles in a
heap-ordered tree (level $l$ holds angles $2^l-1,\ldots,2^{l+1}-2$) and
$2^{n_q}$ phase angles;
a cold start of state $m$ is the basis state $|m\rangle$.  The amplitude
$A_j$ is the product, along the path from the root to leaf $j$, of
$\cos(\theta/2)$ for a left branch and $\sin(\theta/2)$ for a right branch,
and $\varphi_j=-\theta_{z,j}$.  Because a non-root angle rotates only its
subtree, the energy also contains terms linear in $\cos(\theta/2)$ and
$\sin(\theta/2)$, so amplitude coordinates use the shifts $\{0,\pm\pi/2,\pm\pi\}$ and the fit
$a_0+a_1\cos\frac\vartheta2+b_1\sin\frac\vartheta2+a_2\cos\vartheta+b_2\sin\vartheta$ in the shifted angle $\vartheta$, minimized through
the real roots of its stationarity condition; phase coordinates use
$\{0,\pi/2,\pi\}$.  The deflation penalty is $\mu=\|H\|_2$ from at most 30
power iterations on $H^2$ (relative tolerance $10^{-6}$).  After MGSO the eigenvalue of each state is
re-evaluated, and the states are ordered by their Rayleigh quotients.
Gauss--Seidel sweeps update all amplitude angles in tree order and then
the phase angles $1,\ldots,N-1$ (the global phase is fixed), and stop when
the change of the penalized energy between sweeps falls below the
tolerance.  Pure-PW solves are warm-started from the previous step's states (after
an update, from the lifted states); the first bootstrap and contracted
solves and the re-solve after an update start cold.

In each round of the block-coordinate method, the seven amplitude angles
above the blocks are first updated one at a time; the blocks are then
solved from a common snapshot with one local sweep.  Within a block, the
angles, or those of the drawn subset, are taken in tree order in
consecutive groups of four; the four angles of a group are updated
simultaneously from the state reached after the previous group.  At $n_q=4$ each of the eight blocks is a single angle,
and on the three-qubit contracted register the tree is cut one level
higher, giving four one-angle blocks below three top angles; with
randomized subsets, a one-angle block is updated only in the rounds in
which its first subset is drawn.  The Davidson direction is
$r/(\epsilon-{\rm diag}\,H)$ with $r=(H-\epsilon)\psi$, and $|\epsilon-{\rm diag}\,H|$ bounded below by
$\max[10^{-8},10^{-6}(\max|{\rm diag}\,H|+|\epsilon|+1)]$.  The lower eigenstates are projected
out of the Rayleigh--Ritz basis, which is truncated at overlap
eigenvalues below $10^{-10}\max(\lambda_{\max},1)$, and a Ritz update is kept
only if it lowers the penalized energy of the snapshot, re-evaluated in
every round.  The history states of the $n_q=15$ run are the
snapshots of up to eight preceding rounds of the same eigenstate solve.  The four subsets of a block are the angles at positions $s, s+4,
s+8,\ldots$ ($s=0,\ldots,3$) of its tree-ordered angle list; every block
draws its own $s$ in each round from a NumPy~\cite{Harris2020} PCG64 generator with seed 0
that is restarted for every eigenstate solve.  The energy drop is tested
after every round, after a minimum of 16 rounds in the block-parallel
PWOO solves of Tables~\ref{tab:pwoo-main} and~\ref{tab:pwoo-contracted}
(none in the $n_q=15$ run); the block-parallel STO-3G schedules of Tables~\ref{tab:pwoo-shot-scf-gate}
and~\ref{tab:pwoo-contracted} are fixed at 128 rounds.  Round limits and stopping
tolerances are:

\begin{center}
\footnotesize
\begin{tabular}{@{}l l l@{}}
\toprule
Runs & Rotosolve & Rounds, stop \\
\midrule
Table~\ref{tab:main-results}, $n_q=6$ & serial & $\le300$, drop $<10^{-7}$ \\
Table~\ref{tab:main-results}, $n_q=9,12$ & serial & $\le1000$, drop $<10^{-5}$ \\
Table~\ref{tab:pwoo-main} (QC) & block & 16--512, drop $<10^{-8}$ \\
Table~\ref{tab:pwoo-contracted}, cc-pVTZ & block/serial & 16--2000/$\le2000$, $<10^{-8}$$^{a}$ \\
Table~\ref{tab:pwoo-contracted}, STO-3G & block/serial & 128 fixed/$\le2000$, $<10^{-8}$ \\
Table~\ref{tab:pwoo-shot-scf-gate} & block & 128 fixed \\
$n_q=15$, full register & block & $\le1000$, drop $<10^{-6}$ \\
$n_q=15$, contracted & serial & $\le1000$, drop $<10^{-5}$ \\
\bottomrule
\end{tabular}\\[2pt]
$^{a}$Shot row: drop below $\max(10^{-8},3\sigma)$, with $\sigma$ the
propagated shot uncertainty of the drop,
$\sigma=(\sigma_r^2+\sigma_{r-1}^2)^{1/2}$ from the estimated variances of
the two rounds.
\end{center}

\subsection{Shot emulation}

On the full register, the $N_s$ shots of a batch are split equally among
the three blocks of $H_{\rm PWOO}$ (PW--PW, PW--OO plus OO--PW, and
OO--OO); each block estimate is the exact quadratic form
$\langle\psi|O|\psi\rangle$ plus Gaussian noise with variance
$(\langle\psi|\psi\rangle\langle\psi|O^2|\psi\rangle-\langle\psi|O|\psi\rangle^2)$
divided by its number of shots, which also holds for unnormalized vectors.  On the contracted register, Pauli strings with
$|h_j|>10^{-12}$ are grouped greedily in order of decreasing $|h_j|$ into QWC
sets, the shots are split equally among the sets, and counts are sampled
from the multinomial distribution; the estimate is multiplied by
$\langle\psi|\psi\rangle$, so that it estimates $\langle\psi|H|\psi\rangle$ also
for unnormalized vectors.  Each call averages eight batches, and repeated
evaluations of the same state within one SCF step are accumulated.
Rayleigh--Ritz matrix elements are formed by the polarization identity
from shot estimates for $\psi_i\pm\psi_j$ and $\psi_i\pm i\psi_j$, and a
Ritz update is kept only if it lowers the estimated energy.  Shot noise enters only through the energy estimates:
the deflation overlaps, MGSO, the final ordering of the states, and the
reported $E_{\rm KS}$ use the state vectors.  Random numbers come from
NumPy PCG64 generators, seeded with ${\rm seed}+1000003\,t$ at SCF step
$t$ ($+1$ for the contracted-register QWC sampling); two evaluations refer to the same state when their coefficient
vectors agree after removing the phase of the largest component.  The noise-aware criterion compares the mean projector change of the last
five steps with that of the five before, and the quoted energies are
means over the last five steps; the noise-aware runs (separate from the
40-step runs of Fig.~\ref{fig:shot-mechanism}(b)) use $\alpha=0.5$,
seed 1234, and $N_s=10^9$, $10^{10}$, $10^{11}$, and the runs without
randomized subsets $\alpha=0.5$, seed 1234, and $N_s=10^{11}$, $10^{13}$,
$10^{14}$.  The noise-free fixed point of
Table~\ref{tab:pwoo-shot-scf-gate} is a classical SCF with linear mixing,
$\alpha=0.5$, and the single-step test with $|\delta E|<10^{-7}$\,Ha and
$\|\Delta P_{\rm occ}\|_F<10^{-6}$.  Figure~\ref{fig:shot-mechanism}(a) shows seed 1234 at $\alpha=0.5$ (40
steps) and the four $\alpha=0.2$ runs; (b) shows six 40-step $\alpha=0.5$
runs (seed 1234), with the floor the median of the last eight
$\|\Delta P_{\rm occ}\|_F$; the density-matrix change is $\|D_{t+1}-D_t\|_F$.  The fixed-Hamiltonian test of Sec.~\ref{sec:stage5}
uses $H_{\rm KS}$ at the core-Hamiltonian density, eight states, each
started from the corresponding $h_{\rm core}$ eigenvector, randomized
subsets, 128 rounds per state (a 128-round minimum, after which the energy-drop stop at
$10^{-4}$\,Ha applied at once), eight batches of $10^9$ shots per
estimate, and seed 0; the error is that of the exact Rayleigh quotient of
each returned state, and its maximum is taken over all eight states (with
residual directions it occurs on the highest, unoccupied state).

\subsection{Hardware experiment}

The measured PWOO states are those of an earlier STO-3G ($N_{\rm PW}=8$,
$n_g=65$) QC SCF run (block-parallel Rotosolve, linear mixing with
$\alpha=0.5$), stopped when $|\delta E|<10^{-6}$\,Ha and
$\|\Delta P_{\rm occ}\|_F<10^{-4}$ held on a single step (the tolerances of
the three-step PWOO criterion applied on a single step; 21 steps)
($E_{\rm KS}=-4.196939$, $9\,\mu$Ha above the Table~\ref{tab:pwoo-main}
value), and the AO-only states those of the classical AO-only SCF
($-4.189879$); the exact difference for these states is $-7.06$\,mHa.
The observables are $H_{\rm KS}$ built from the final density matrix of
each source calculation and, on the contracted registers,
$\tilde H=U^\dagger H_{\rm KS}U$ with $U$ the orthonormal QR factor of the eight stored
orbitals in ascending-eigenvalue order, decomposed into
Pauli strings and grouped into QWC sets as in the shot emulation.
Circuits are built with Qiskit's state preparation~\cite{Qiskit2024,Iten2016}
from the stored amplitudes and transpiled (Qiskit 2.4, qiskit-ionq
1.0.2~\cite{QiskitIonQ}, optimization level 1) for \texttt{ionq\_qpu.forte-1} with debiasing enabled and results retrieved without the sharpening
option (the provider's default aggregation); native compilation is done
by the provider.  The two-qubit gate counts are CNOT gates of Qiskit's
standard gate set after transpilation, before the provider's native
compilation: the four contracted-register circuits have 0, 4, 1, and 4
(PWOO) and 4, 3, 1, and 4 (AO-only), and the full-register circuits 11
(PWOO) and 4 (AO-only).  On the contracted registers the four occupied states are measured with 4096 shots
per QWC circuit (27 sets for PWOO, 5 for AO-only).  On the full registers,
the lowest orbital is measured with 81 (PWOO) and 5 (AO-only) QWC sets and shots
proportional to each set's standard deviation, with at least 500 per
set ($5.51\times10^5$ and $5.40\times10^5$ shots per repetition).  The
uncertainty of the contracted-register difference is the sample standard
deviation ($n-1$ normalization) of the two paired repetitions; the
full-register $\pm2.7$\,mHa combines the two systems' standard
deviations over the repetitions in quadrature, and the calibrated difference carries the propagated shot and
fit uncertainty of $\lambda$.  The depolarizing model is
$\langle O_g\rangle_{\rm meas}=\lambda\langle O_g\rangle_{\rm exact}$ (identity
components removed), fit by weighted least squares with weights
$N_g/\sigma_g^2$, where $N_g$ is the number of shots of set $g$ and
$\sigma_g$ its exact single-shot standard deviation; the calibrated eigenvalue is
$h_I+(\epsilon^{\rm meas}-h_I)/\lambda$, with $h_I$ the identity
coefficient.  The quoted $\lambda$ is the mean of the per-repetition fits;
each fit's uncertainty is scaled by $\sqrt{\chi^2/{\rm dof}}$ when
$\chi^2/{\rm dof}>1$.  The experiment ran 6--9 July 2026 in 598 jobs (circuits are split into
jobs of at most $10^4$ shots), whose identifiers are archived.

\bibliography{references}

\begin{thebibliography}{69}%
\makeatletter
\providecommand \@ifxundefined [1]{%
 \@ifx{#1\undefined}
}%
\providecommand \@ifnum [1]{%
 \ifnum #1\expandafter \@firstoftwo
 \else \expandafter \@secondoftwo
 \fi
}%
\providecommand \@ifx [1]{%
 \ifx #1\expandafter \@firstoftwo
 \else \expandafter \@secondoftwo
 \fi
}%
\providecommand \natexlab [1]{#1}%
\providecommand \enquote  [1]{``#1''}%
\providecommand \bibnamefont  [1]{#1}%
\providecommand \bibfnamefont [1]{#1}%
\providecommand \citenamefont [1]{#1}%
\providecommand \href@noop [0]{\@secondoftwo}%
\providecommand \href [0]{\begingroup \@sanitize@url \@href}%
\providecommand \@href[1]{\@@startlink{#1}\@@href}%
\providecommand \@@href[1]{\endgroup#1\@@endlink}%
\providecommand \@sanitize@url [0]{\catcode `\\12\catcode `\$12\catcode
  `\&12\catcode `\#12\catcode `\^12\catcode `\_12\catcode `\%12\relax}%
\providecommand \@@startlink[1]{}%
\providecommand \@@endlink[0]{}%
\providecommand \url  [0]{\begingroup\@sanitize@url \@url }%
\providecommand \@url [1]{\endgroup\@href {#1}{\urlprefix }}%
\providecommand \urlprefix  [0]{URL }%
\providecommand \Eprint [0]{\href }%
\providecommand \doibase [0]{https://doi.org/}%
\providecommand \selectlanguage [0]{\@gobble}%
\providecommand \bibinfo  [0]{\@secondoftwo}%
\providecommand \bibfield  [0]{\@secondoftwo}%
\providecommand \translation [1]{[#1]}%
\providecommand \BibitemOpen [0]{}%
\providecommand \bibitemStop [0]{}%
\providecommand \bibitemNoStop [0]{.\EOS\space}%
\providecommand \EOS [0]{\spacefactor3000\relax}%
\providecommand \BibitemShut  [1]{\csname bibitem#1\endcsname}%
\let\auto@bib@innerbib\@empty
\bibitem [{\citenamefont {Hohenberg}\ and\ \citenamefont
  {Kohn}(1964)}]{Hohenberg1964}%
  \BibitemOpen
  \bibfield  {author} {\bibinfo {author} {\bibfnamefont {P.}~\bibnamefont
  {Hohenberg}}\ and\ \bibinfo {author} {\bibfnamefont {W.}~\bibnamefont
  {Kohn}},\ }\bibfield  {title} {\bibinfo {title} {Inhomogeneous electron
  gas},\ }\href {https://doi.org/10.1103/PhysRev.136.B864} {\bibfield
  {journal} {\bibinfo  {journal} {Phys. Rev.}\ }\textbf {\bibinfo {volume}
  {136}},\ \bibinfo {pages} {B864} (\bibinfo {year} {1964})}\BibitemShut
  {NoStop}%
\bibitem [{\citenamefont {Kohn}\ and\ \citenamefont {Sham}(1965)}]{Kohn1965}%
  \BibitemOpen
  \bibfield  {author} {\bibinfo {author} {\bibfnamefont {W.}~\bibnamefont
  {Kohn}}\ and\ \bibinfo {author} {\bibfnamefont {L.~J.}\ \bibnamefont
  {Sham}},\ }\bibfield  {title} {\bibinfo {title} {Self-consistent equations
  including exchange and correlation effects},\ }\href
  {https://doi.org/10.1103/PhysRev.140.A1133} {\bibfield  {journal} {\bibinfo
  {journal} {Phys. Rev.}\ }\textbf {\bibinfo {volume} {140}},\ \bibinfo {pages}
  {A1133} (\bibinfo {year} {1965})}\BibitemShut {NoStop}%
\bibitem [{\citenamefont {Becke}(1993)}]{Becke1993}%
  \BibitemOpen
  \bibfield  {author} {\bibinfo {author} {\bibfnamefont {A.~D.}\ \bibnamefont
  {Becke}},\ }\bibfield  {title} {\bibinfo {title} {Density-functional
  thermochemistry. {III}. {T}he role of exact exchange},\ }\href
  {https://doi.org/10.1063/1.464913} {\bibfield  {journal} {\bibinfo  {journal}
  {J. Chem. Phys.}\ }\textbf {\bibinfo {volume} {98}},\ \bibinfo {pages} {5648}
  (\bibinfo {year} {1993})}\BibitemShut {NoStop}%
\bibitem [{\citenamefont {Anisimov}\ \emph {et~al.}(1991)\citenamefont
  {Anisimov}, \citenamefont {Zaanen},\ and\ \citenamefont
  {Andersen}}]{Anisimov1991}%
  \BibitemOpen
  \bibfield  {author} {\bibinfo {author} {\bibfnamefont {V.~I.}\ \bibnamefont
  {Anisimov}}, \bibinfo {author} {\bibfnamefont {J.}~\bibnamefont {Zaanen}},\
  and\ \bibinfo {author} {\bibfnamefont {O.~K.}\ \bibnamefont {Andersen}},\
  }\bibfield  {title} {\bibinfo {title} {Band theory and {Mott} insulators:
  {Hubbard} {$U$} instead of {Stoner} {$I$}},\ }\href
  {https://doi.org/10.1103/PhysRevB.44.943} {\bibfield  {journal} {\bibinfo
  {journal} {Phys. Rev. B}\ }\textbf {\bibinfo {volume} {44}},\ \bibinfo
  {pages} {943} (\bibinfo {year} {1991})}\BibitemShut {NoStop}%
\bibitem [{\citenamefont {Kotliar}\ \emph {et~al.}(2006)\citenamefont
  {Kotliar}, \citenamefont {Savrasov}, \citenamefont {Haule}, \citenamefont
  {Oudovenko}, \citenamefont {Parcollet},\ and\ \citenamefont
  {Marianetti}}]{Kotliar2006}%
  \BibitemOpen
  \bibfield  {author} {\bibinfo {author} {\bibfnamefont {G.}~\bibnamefont
  {Kotliar}}, \bibinfo {author} {\bibfnamefont {S.~Y.}\ \bibnamefont
  {Savrasov}}, \bibinfo {author} {\bibfnamefont {K.}~\bibnamefont {Haule}},
  \bibinfo {author} {\bibfnamefont {V.~S.}\ \bibnamefont {Oudovenko}}, \bibinfo
  {author} {\bibfnamefont {O.}~\bibnamefont {Parcollet}},\ and\ \bibinfo
  {author} {\bibfnamefont {C.~A.}\ \bibnamefont {Marianetti}},\ }\bibfield
  {title} {\bibinfo {title} {Electronic structure calculations with dynamical
  mean-field theory},\ }\href {https://doi.org/10.1103/RevModPhys.78.865}
  {\bibfield  {journal} {\bibinfo  {journal} {Rev. Mod. Phys.}\ }\textbf
  {\bibinfo {volume} {78}},\ \bibinfo {pages} {865} (\bibinfo {year}
  {2006})}\BibitemShut {NoStop}%
\bibitem [{\citenamefont {McClean}\ \emph {et~al.}(2016)\citenamefont
  {McClean}, \citenamefont {Romero}, \citenamefont {Babbush},\ and\
  \citenamefont {Aspuru-Guzik}}]{McClean2016}%
  \BibitemOpen
  \bibfield  {author} {\bibinfo {author} {\bibfnamefont {J.~R.}\ \bibnamefont
  {McClean}}, \bibinfo {author} {\bibfnamefont {J.}~\bibnamefont {Romero}},
  \bibinfo {author} {\bibfnamefont {R.}~\bibnamefont {Babbush}},\ and\ \bibinfo
  {author} {\bibfnamefont {A.}~\bibnamefont {Aspuru-Guzik}},\ }\bibfield
  {title} {\bibinfo {title} {The theory of variational hybrid quantum-classical
  algorithms},\ }\href {https://doi.org/10.1088/1367-2630/18/2/023023}
  {\bibfield  {journal} {\bibinfo  {journal} {New J. Phys.}\ }\textbf {\bibinfo
  {volume} {18}},\ \bibinfo {pages} {023023} (\bibinfo {year} {2016})},\
  \Eprint {https://arxiv.org/abs/1509.04279} {arXiv:1509.04279} \BibitemShut
  {NoStop}%
\bibitem [{\citenamefont {Bauer}\ \emph {et~al.}(2020)\citenamefont {Bauer},
  \citenamefont {Bravyi}, \citenamefont {Motta},\ and\ \citenamefont
  {Chan}}]{Bauer2020}%
  \BibitemOpen
  \bibfield  {author} {\bibinfo {author} {\bibfnamefont {B.}~\bibnamefont
  {Bauer}}, \bibinfo {author} {\bibfnamefont {S.}~\bibnamefont {Bravyi}},
  \bibinfo {author} {\bibfnamefont {M.}~\bibnamefont {Motta}},\ and\ \bibinfo
  {author} {\bibfnamefont {G.~K.-L.}\ \bibnamefont {Chan}},\ }\bibfield
  {title} {\bibinfo {title} {Quantum algorithms for quantum chemistry and
  quantum materials science},\ }\href
  {https://doi.org/10.1021/acs.chemrev.9b00829} {\bibfield  {journal} {\bibinfo
   {journal} {Chem. Rev.}\ }\textbf {\bibinfo {volume} {120}},\ \bibinfo
  {pages} {12685} (\bibinfo {year} {2020})}\BibitemShut {NoStop}%
\bibitem [{\citenamefont {McArdle}\ \emph {et~al.}(2020)\citenamefont
  {McArdle}, \citenamefont {Endo}, \citenamefont {Aspuru-Guzik}, \citenamefont
  {Benjamin},\ and\ \citenamefont {Yuan}}]{McArdle2020}%
  \BibitemOpen
  \bibfield  {author} {\bibinfo {author} {\bibfnamefont {S.}~\bibnamefont
  {McArdle}}, \bibinfo {author} {\bibfnamefont {S.}~\bibnamefont {Endo}},
  \bibinfo {author} {\bibfnamefont {A.}~\bibnamefont {Aspuru-Guzik}}, \bibinfo
  {author} {\bibfnamefont {S.~C.}\ \bibnamefont {Benjamin}},\ and\ \bibinfo
  {author} {\bibfnamefont {X.}~\bibnamefont {Yuan}},\ }\bibfield  {title}
  {\bibinfo {title} {Quantum computational chemistry},\ }\href
  {https://doi.org/10.1103/RevModPhys.92.015003} {\bibfield  {journal}
  {\bibinfo  {journal} {Rev. Mod. Phys.}\ }\textbf {\bibinfo {volume} {92}},\
  \bibinfo {pages} {015003} (\bibinfo {year} {2020})}\BibitemShut {NoStop}%
\bibitem [{\citenamefont {Babbush}\ \emph {et~al.}(2018)\citenamefont
  {Babbush}, \citenamefont {Wiebe}, \citenamefont {McClean}, \citenamefont
  {McClain}, \citenamefont {Neven},\ and\ \citenamefont {Chan}}]{Babbush2018}%
  \BibitemOpen
  \bibfield  {author} {\bibinfo {author} {\bibfnamefont {R.}~\bibnamefont
  {Babbush}}, \bibinfo {author} {\bibfnamefont {N.}~\bibnamefont {Wiebe}},
  \bibinfo {author} {\bibfnamefont {J.}~\bibnamefont {McClean}}, \bibinfo
  {author} {\bibfnamefont {J.}~\bibnamefont {McClain}}, \bibinfo {author}
  {\bibfnamefont {H.}~\bibnamefont {Neven}},\ and\ \bibinfo {author}
  {\bibfnamefont {G.~K.-L.}\ \bibnamefont {Chan}},\ }\bibfield  {title}
  {\bibinfo {title} {Low-depth quantum simulation of materials},\ }\href
  {https://doi.org/10.1103/PhysRevX.8.011044} {\bibfield  {journal} {\bibinfo
  {journal} {Phys. Rev. X}\ }\textbf {\bibinfo {volume} {8}},\ \bibinfo {pages}
  {011044} (\bibinfo {year} {2018})},\ \Eprint
  {https://arxiv.org/abs/1706.00023} {arXiv:1706.00023} \BibitemShut {NoStop}%
\bibitem [{\citenamefont {Babbush}\ \emph {et~al.}(2019)\citenamefont
  {Babbush}, \citenamefont {Berry}, \citenamefont {McClean},\ and\
  \citenamefont {Neven}}]{Babbush2019}%
  \BibitemOpen
  \bibfield  {author} {\bibinfo {author} {\bibfnamefont {R.}~\bibnamefont
  {Babbush}}, \bibinfo {author} {\bibfnamefont {D.~W.}\ \bibnamefont {Berry}},
  \bibinfo {author} {\bibfnamefont {J.~R.}\ \bibnamefont {McClean}},\ and\
  \bibinfo {author} {\bibfnamefont {H.}~\bibnamefont {Neven}},\ }\bibfield
  {title} {\bibinfo {title} {Quantum simulation of chemistry with sublinear
  scaling in basis size},\ }\href {https://doi.org/10.1038/s41534-019-0199-y}
  {\bibfield  {journal} {\bibinfo  {journal} {npj Quantum Inf.}\ }\textbf
  {\bibinfo {volume} {5}},\ \bibinfo {pages} {92} (\bibinfo {year} {2019})},\
  \Eprint {https://arxiv.org/abs/1807.09802} {arXiv:1807.09802} \BibitemShut
  {NoStop}%
\bibitem [{\citenamefont {Su}\ \emph {et~al.}(2021)\citenamefont {Su},
  \citenamefont {Berry}, \citenamefont {Wiebe}, \citenamefont {Rubin},\ and\
  \citenamefont {Babbush}}]{Su2021}%
  \BibitemOpen
  \bibfield  {author} {\bibinfo {author} {\bibfnamefont {Y.}~\bibnamefont
  {Su}}, \bibinfo {author} {\bibfnamefont {D.~W.}\ \bibnamefont {Berry}},
  \bibinfo {author} {\bibfnamefont {N.}~\bibnamefont {Wiebe}}, \bibinfo
  {author} {\bibfnamefont {N.}~\bibnamefont {Rubin}},\ and\ \bibinfo {author}
  {\bibfnamefont {R.}~\bibnamefont {Babbush}},\ }\bibfield  {title} {\bibinfo
  {title} {Fault-tolerant quantum simulations of chemistry in first
  quantization},\ }\href {https://doi.org/10.1103/PRXQuantum.2.040332}
  {\bibfield  {journal} {\bibinfo  {journal} {PRX Quantum}\ }\textbf {\bibinfo
  {volume} {2}},\ \bibinfo {pages} {040332} (\bibinfo {year} {2021})},\ \Eprint
  {https://arxiv.org/abs/2105.12767} {arXiv:2105.12767} \BibitemShut {NoStop}%
\bibitem [{\citenamefont {Berry}\ \emph {et~al.}(2024)\citenamefont {Berry},
  \citenamefont {Rubin}, \citenamefont {Elnabawy}, \citenamefont {Ahlers},
  \citenamefont {DePrince}, \citenamefont {Lee}, \citenamefont {Gogolin},\ and\
  \citenamefont {Babbush}}]{Berry2024}%
  \BibitemOpen
  \bibfield  {author} {\bibinfo {author} {\bibfnamefont {D.~W.}\ \bibnamefont
  {Berry}}, \bibinfo {author} {\bibfnamefont {N.~C.}\ \bibnamefont {Rubin}},
  \bibinfo {author} {\bibfnamefont {A.~O.}\ \bibnamefont {Elnabawy}}, \bibinfo
  {author} {\bibfnamefont {G.}~\bibnamefont {Ahlers}}, \bibinfo {author}
  {\bibfnamefont {A.~E.}\ \bibnamefont {DePrince}, \bibfnamefont {III}},
  \bibinfo {author} {\bibfnamefont {J.}~\bibnamefont {Lee}}, \bibinfo {author}
  {\bibfnamefont {C.}~\bibnamefont {Gogolin}},\ and\ \bibinfo {author}
  {\bibfnamefont {R.}~\bibnamefont {Babbush}},\ }\bibfield  {title} {\bibinfo
  {title} {Quantum simulation of realistic materials in first quantization
  using non-local pseudopotentials},\ }\href
  {https://doi.org/10.1038/s41534-024-00896-9} {\bibfield  {journal} {\bibinfo
  {journal} {npj Quantum Information}\ }\textbf {\bibinfo {volume} {10}},\
  \bibinfo {pages} {130} (\bibinfo {year} {2024})},\ \Eprint
  {https://arxiv.org/abs/2312.07654} {arXiv:2312.07654} \BibitemShut {NoStop}%
\bibitem [{\citenamefont {Georges}\ \emph {et~al.}(2025)\citenamefont
  {Georges}, \citenamefont {Bothe}, \citenamefont {S{\"u}nderhauf},
  \citenamefont {Berntson}, \citenamefont {Izs{\'a}k},\ and\ \citenamefont
  {Ivanov}}]{Georges2025}%
  \BibitemOpen
  \bibfield  {author} {\bibinfo {author} {\bibfnamefont {T.~N.}\ \bibnamefont
  {Georges}}, \bibinfo {author} {\bibfnamefont {M.}~\bibnamefont {Bothe}},
  \bibinfo {author} {\bibfnamefont {C.}~\bibnamefont {S{\"u}nderhauf}},
  \bibinfo {author} {\bibfnamefont {B.~K.}\ \bibnamefont {Berntson}}, \bibinfo
  {author} {\bibfnamefont {R.}~\bibnamefont {Izs{\'a}k}},\ and\ \bibinfo
  {author} {\bibfnamefont {A.~V.}\ \bibnamefont {Ivanov}},\ }\bibfield  {title}
  {\bibinfo {title} {Quantum simulations of chemistry in first quantization
  with any basis set},\ }\href {https://doi.org/10.1038/s41534-025-00987-1}
  {\bibfield  {journal} {\bibinfo  {journal} {npj Quantum Inf.}\ }\textbf
  {\bibinfo {volume} {11}},\ \bibinfo {pages} {55} (\bibinfo {year} {2025})},\
  \Eprint {https://arxiv.org/abs/2408.03145} {arXiv:2408.03145} \BibitemShut
  {NoStop}%
\bibitem [{\citenamefont {Ivanov}\ \emph {et~al.}(2025)\citenamefont {Ivanov},
  \citenamefont {Patterson}, \citenamefont {Bothe}, \citenamefont
  {S{\"u}nderhauf}, \citenamefont {Berntson}, \citenamefont {Mortensen},
  \citenamefont {Kuisma}, \citenamefont {Campbell},\ and\ \citenamefont
  {Izs{\'a}k}}]{Ivanov2025}%
  \BibitemOpen
  \bibfield  {author} {\bibinfo {author} {\bibfnamefont {A.~V.}\ \bibnamefont
  {Ivanov}}, \bibinfo {author} {\bibfnamefont {A.}~\bibnamefont {Patterson}},
  \bibinfo {author} {\bibfnamefont {M.}~\bibnamefont {Bothe}}, \bibinfo
  {author} {\bibfnamefont {C.}~\bibnamefont {S{\"u}nderhauf}}, \bibinfo
  {author} {\bibfnamefont {B.~K.}\ \bibnamefont {Berntson}}, \bibinfo {author}
  {\bibfnamefont {J.~J.}\ \bibnamefont {Mortensen}}, \bibinfo {author}
  {\bibfnamefont {M.}~\bibnamefont {Kuisma}}, \bibinfo {author} {\bibfnamefont
  {E.}~\bibnamefont {Campbell}},\ and\ \bibinfo {author} {\bibfnamefont
  {R.}~\bibnamefont {Izs{\'a}k}},\ }\bibfield  {title} {\bibinfo {title}
  {Quantum computation of electronic structure with projector augmented-wave
  method and plane wave basis set},\ }\href
  {https://doi.org/10.1021/acs.jctc.5c00551} {\bibfield  {journal} {\bibinfo
  {journal} {J. Chem. Theory Comput.}\ }\textbf {\bibinfo {volume} {21}},\
  \bibinfo {pages} {7360} (\bibinfo {year} {2025})},\ \Eprint
  {https://arxiv.org/abs/2408.03159} {arXiv:2408.03159} \BibitemShut {NoStop}%
\bibitem [{\citenamefont {Hatcher}\ \emph {et~al.}(2019)\citenamefont
  {Hatcher}, \citenamefont {Kittl},\ and\ \citenamefont {Bowen}}]{Hatcher2019}%
  \BibitemOpen
  \bibfield  {author} {\bibinfo {author} {\bibfnamefont {R.}~\bibnamefont
  {Hatcher}}, \bibinfo {author} {\bibfnamefont {J.~A.}\ \bibnamefont {Kittl}},\
  and\ \bibinfo {author} {\bibfnamefont {C.}~\bibnamefont {Bowen}},\
  }\href@noop {} {\bibinfo {title} {A method to calculate correlation for
  density functional theory on a quantum processor}} (\bibinfo {year} {2019}),\
  \Eprint {https://arxiv.org/abs/1903.05550} {arXiv:1903.05550} \BibitemShut
  {NoStop}%
\bibitem [{\citenamefont {Baker}\ and\ \citenamefont
  {Poulin}(2020)}]{Baker2020}%
  \BibitemOpen
  \bibfield  {author} {\bibinfo {author} {\bibfnamefont {T.~E.}\ \bibnamefont
  {Baker}}\ and\ \bibinfo {author} {\bibfnamefont {D.}~\bibnamefont {Poulin}},\
  }\bibfield  {title} {\bibinfo {title} {Density functionals and {K}ohn--{S}ham
  potentials with minimal wavefunction preparations on a quantum computer},\
  }\href {https://doi.org/10.1103/PhysRevResearch.2.043238} {\bibfield
  {journal} {\bibinfo  {journal} {Phys. Rev. Research}\ }\textbf {\bibinfo
  {volume} {2}},\ \bibinfo {pages} {043238} (\bibinfo {year} {2020})},\ \Eprint
  {https://arxiv.org/abs/2008.05592} {arXiv:2008.05592} \BibitemShut {NoStop}%
\bibitem [{\citenamefont {Sheridan}\ \emph {et~al.}(2024)\citenamefont
  {Sheridan}, \citenamefont {Mineh}, \citenamefont {Santos},\ and\
  \citenamefont {Cubitt}}]{Sheridan2024}%
  \BibitemOpen
  \bibfield  {author} {\bibinfo {author} {\bibfnamefont {E.}~\bibnamefont
  {Sheridan}}, \bibinfo {author} {\bibfnamefont {L.}~\bibnamefont {Mineh}},
  \bibinfo {author} {\bibfnamefont {R.~A.}\ \bibnamefont {Santos}},\ and\
  \bibinfo {author} {\bibfnamefont {T.}~\bibnamefont {Cubitt}},\ }\href@noop {}
  {\bibinfo {title} {Enhancing density functional theory using the variational
  quantum eigensolver}} (\bibinfo {year} {2024}),\ \Eprint
  {https://arxiv.org/abs/2402.18534} {arXiv:2402.18534} \BibitemShut {NoStop}%
\bibitem [{\citenamefont {Cerasoli}\ \emph {et~al.}(2020)\citenamefont
  {Cerasoli}, \citenamefont {Sherbert}, \citenamefont {S{\l}awi{\'n}ska},\ and\
  \citenamefont {Buongiorno~Nardelli}}]{Cerasoli2020}%
  \BibitemOpen
  \bibfield  {author} {\bibinfo {author} {\bibfnamefont {F.~T.}\ \bibnamefont
  {Cerasoli}}, \bibinfo {author} {\bibfnamefont {K.}~\bibnamefont {Sherbert}},
  \bibinfo {author} {\bibfnamefont {J.}~\bibnamefont {S{\l}awi{\'n}ska}},\ and\
  \bibinfo {author} {\bibfnamefont {M.}~\bibnamefont {Buongiorno~Nardelli}},\
  }\bibfield  {title} {\bibinfo {title} {Quantum computation of silicon
  electronic band structure},\ }\href {https://doi.org/10.1039/D0CP04008H}
  {\bibfield  {journal} {\bibinfo  {journal} {Phys. Chem. Chem. Phys.}\
  }\textbf {\bibinfo {volume} {22}},\ \bibinfo {pages} {21816} (\bibinfo {year}
  {2020})},\ \Eprint {https://arxiv.org/abs/2006.03807} {arXiv:2006.03807}
  \BibitemShut {NoStop}%
\bibitem [{\citenamefont {Sherbert}\ \emph {et~al.}(2021)\citenamefont
  {Sherbert}, \citenamefont {Cerasoli},\ and\ \citenamefont
  {Buongiorno~Nardelli}}]{Sherbert2021}%
  \BibitemOpen
  \bibfield  {author} {\bibinfo {author} {\bibfnamefont {K.}~\bibnamefont
  {Sherbert}}, \bibinfo {author} {\bibfnamefont {F.}~\bibnamefont {Cerasoli}},\
  and\ \bibinfo {author} {\bibfnamefont {M.}~\bibnamefont
  {Buongiorno~Nardelli}},\ }\bibfield  {title} {\bibinfo {title} {A systematic
  variational approach to band theory in a quantum computer},\ }\href
  {https://doi.org/10.1039/D1RA07451B} {\bibfield  {journal} {\bibinfo
  {journal} {RSC Adv.}\ }\textbf {\bibinfo {volume} {11}},\ \bibinfo {pages}
  {39438} (\bibinfo {year} {2021})},\ \Eprint
  {https://arxiv.org/abs/2104.03409} {arXiv:2104.03409} \BibitemShut {NoStop}%
\bibitem [{\citenamefont {Ko}\ \emph {et~al.}(2023)\citenamefont {Ko},
  \citenamefont {Li},\ and\ \citenamefont {Wang}}]{Ko2023}%
  \BibitemOpen
  \bibfield  {author} {\bibinfo {author} {\bibfnamefont {T.}~\bibnamefont
  {Ko}}, \bibinfo {author} {\bibfnamefont {X.}~\bibnamefont {Li}},\ and\
  \bibinfo {author} {\bibfnamefont {C.}~\bibnamefont {Wang}},\ }\href@noop {}
  {\bibinfo {title} {Implementation of the density-functional theory on quantum
  computers with linear scaling with respect to the number of atoms}} (\bibinfo
  {year} {2023}),\ \Eprint {https://arxiv.org/abs/2307.07067}
  {arXiv:2307.07067} \BibitemShut {NoStop}%
\bibitem [{\citenamefont {Senjean}\ \emph {et~al.}(2023)\citenamefont
  {Senjean}, \citenamefont {Yalouz},\ and\ \citenamefont
  {Sauban{\`e}re}}]{Senjean2022}%
  \BibitemOpen
  \bibfield  {author} {\bibinfo {author} {\bibfnamefont {B.}~\bibnamefont
  {Senjean}}, \bibinfo {author} {\bibfnamefont {S.}~\bibnamefont {Yalouz}},\
  and\ \bibinfo {author} {\bibfnamefont {M.}~\bibnamefont {Sauban{\`e}re}},\
  }\bibfield  {title} {\bibinfo {title} {Toward density functional theory on
  quantum computers?},\ }\href {https://doi.org/10.21468/SciPostPhys.14.3.055}
  {\bibfield  {journal} {\bibinfo  {journal} {SciPost Phys.}\ }\textbf
  {\bibinfo {volume} {14}},\ \bibinfo {pages} {055} (\bibinfo {year} {2023})},\
  \Eprint {https://arxiv.org/abs/2204.01443} {arXiv:2204.01443} \BibitemShut
  {NoStop}%
\bibitem [{\citenamefont {Slater}(1951)}]{Slater1951}%
  \BibitemOpen
  \bibfield  {author} {\bibinfo {author} {\bibfnamefont {J.~C.}\ \bibnamefont
  {Slater}},\ }\bibfield  {title} {\bibinfo {title} {A simplification of the
  {Hartree-Fock} method},\ }\href {https://doi.org/10.1103/PhysRev.81.385}
  {\bibfield  {journal} {\bibinfo  {journal} {Phys. Rev.}\ }\textbf {\bibinfo
  {volume} {81}},\ \bibinfo {pages} {385} (\bibinfo {year} {1951})}\BibitemShut
  {NoStop}%
\bibitem [{\citenamefont {Vosko}\ \emph {et~al.}(1980)\citenamefont {Vosko},
  \citenamefont {Wilk},\ and\ \citenamefont {Nusair}}]{Vosko1980}%
  \BibitemOpen
  \bibfield  {author} {\bibinfo {author} {\bibfnamefont {S.~H.}\ \bibnamefont
  {Vosko}}, \bibinfo {author} {\bibfnamefont {L.}~\bibnamefont {Wilk}},\ and\
  \bibinfo {author} {\bibfnamefont {M.}~\bibnamefont {Nusair}},\ }\bibfield
  {title} {\bibinfo {title} {Accurate spin-dependent electron liquid
  correlation energies for local spin density calculations: a critical
  analysis},\ }\href {https://doi.org/10.1139/p80-159} {\bibfield  {journal}
  {\bibinfo  {journal} {Can. J. Phys.}\ }\textbf {\bibinfo {volume} {58}},\
  \bibinfo {pages} {1200} (\bibinfo {year} {1980})}\BibitemShut {NoStop}%
\bibitem [{\citenamefont {Hehre}\ \emph {et~al.}(1969)\citenamefont {Hehre},
  \citenamefont {Stewart},\ and\ \citenamefont {Pople}}]{Hehre1969}%
  \BibitemOpen
  \bibfield  {author} {\bibinfo {author} {\bibfnamefont {W.~J.}\ \bibnamefont
  {Hehre}}, \bibinfo {author} {\bibfnamefont {R.~F.}\ \bibnamefont {Stewart}},\
  and\ \bibinfo {author} {\bibfnamefont {J.~A.}\ \bibnamefont {Pople}},\
  }\bibfield  {title} {\bibinfo {title} {Self-consistent molecular-orbital
  methods. {I}. {U}se of {G}aussian expansions of {S}later-type atomic
  orbitals},\ }\href {https://doi.org/10.1063/1.1672392} {\bibfield  {journal}
  {\bibinfo  {journal} {J. Chem. Phys.}\ }\textbf {\bibinfo {volume} {51}},\
  \bibinfo {pages} {2657} (\bibinfo {year} {1969})}\BibitemShut {NoStop}%
\bibitem [{\citenamefont {Nakanishi}\ \emph {et~al.}(2019)\citenamefont
  {Nakanishi}, \citenamefont {Mitarai},\ and\ \citenamefont
  {Fujii}}]{Nakanishi2019}%
  \BibitemOpen
  \bibfield  {author} {\bibinfo {author} {\bibfnamefont {K.~M.}\ \bibnamefont
  {Nakanishi}}, \bibinfo {author} {\bibfnamefont {K.}~\bibnamefont {Mitarai}},\
  and\ \bibinfo {author} {\bibfnamefont {K.}~\bibnamefont {Fujii}},\ }\bibfield
   {title} {\bibinfo {title} {Subspace-search variational quantum eigensolver
  for excited states},\ }\href
  {https://doi.org/10.1103/PhysRevResearch.1.033062} {\bibfield  {journal}
  {\bibinfo  {journal} {Phys. Rev. Research}\ }\textbf {\bibinfo {volume}
  {1}},\ \bibinfo {pages} {033062} (\bibinfo {year} {2019})},\ \Eprint
  {https://arxiv.org/abs/1810.09434} {arXiv:1810.09434} \BibitemShut {NoStop}%
\bibitem [{\citenamefont {Nakanishi}\ \emph {et~al.}(2020)\citenamefont
  {Nakanishi}, \citenamefont {Fujii},\ and\ \citenamefont
  {Todo}}]{Nakanishi2020}%
  \BibitemOpen
  \bibfield  {author} {\bibinfo {author} {\bibfnamefont {K.~M.}\ \bibnamefont
  {Nakanishi}}, \bibinfo {author} {\bibfnamefont {K.}~\bibnamefont {Fujii}},\
  and\ \bibinfo {author} {\bibfnamefont {S.}~\bibnamefont {Todo}},\ }\bibfield
  {title} {\bibinfo {title} {Sequential minimal optimization for
  quantum-classical hybrid algorithms},\ }\href
  {https://doi.org/10.1103/PhysRevResearch.2.043158} {\bibfield  {journal}
  {\bibinfo  {journal} {Phys. Rev. Research}\ }\textbf {\bibinfo {volume}
  {2}},\ \bibinfo {pages} {043158} (\bibinfo {year} {2020})},\ \Eprint
  {https://arxiv.org/abs/1903.12166} {arXiv:1903.12166} \BibitemShut {NoStop}%
\bibitem [{\citenamefont {Ostaszewski}\ \emph {et~al.}(2021)\citenamefont
  {Ostaszewski}, \citenamefont {Grant},\ and\ \citenamefont
  {Benedetti}}]{Ostaszewski2021}%
  \BibitemOpen
  \bibfield  {author} {\bibinfo {author} {\bibfnamefont {M.}~\bibnamefont
  {Ostaszewski}}, \bibinfo {author} {\bibfnamefont {E.}~\bibnamefont {Grant}},\
  and\ \bibinfo {author} {\bibfnamefont {M.}~\bibnamefont {Benedetti}},\
  }\bibfield  {title} {\bibinfo {title} {Structure optimization for
  parameterized quantum circuits},\ }\href
  {https://doi.org/10.22331/q-2021-01-28-391} {\bibfield  {journal} {\bibinfo
  {journal} {Quantum}\ }\textbf {\bibinfo {volume} {5}},\ \bibinfo {pages}
  {391} (\bibinfo {year} {2021})},\ \Eprint {https://arxiv.org/abs/1905.09692}
  {arXiv:1905.09692} \BibitemShut {NoStop}%
\bibitem [{\citenamefont {Wierichs}\ \emph {et~al.}(2022)\citenamefont
  {Wierichs}, \citenamefont {Izaac}, \citenamefont {Wang},\ and\ \citenamefont
  {Lin}}]{Wierichs2022}%
  \BibitemOpen
  \bibfield  {author} {\bibinfo {author} {\bibfnamefont {D.}~\bibnamefont
  {Wierichs}}, \bibinfo {author} {\bibfnamefont {J.}~\bibnamefont {Izaac}},
  \bibinfo {author} {\bibfnamefont {C.}~\bibnamefont {Wang}},\ and\ \bibinfo
  {author} {\bibfnamefont {C.~Y.-Y.}\ \bibnamefont {Lin}},\ }\bibfield  {title}
  {\bibinfo {title} {General parameter-shift rules for quantum gradients},\
  }\href {https://doi.org/10.22331/q-2022-03-30-677} {\bibfield  {journal}
  {\bibinfo  {journal} {Quantum}\ }\textbf {\bibinfo {volume} {6}},\ \bibinfo
  {pages} {677} (\bibinfo {year} {2022})},\ \Eprint
  {https://arxiv.org/abs/2107.12390} {arXiv:2107.12390} \BibitemShut {NoStop}%
\bibitem [{\citenamefont {Higgott}\ \emph {et~al.}(2019)\citenamefont
  {Higgott}, \citenamefont {Wang},\ and\ \citenamefont
  {Brierley}}]{Higgott2019}%
  \BibitemOpen
  \bibfield  {author} {\bibinfo {author} {\bibfnamefont {O.}~\bibnamefont
  {Higgott}}, \bibinfo {author} {\bibfnamefont {D.}~\bibnamefont {Wang}},\ and\
  \bibinfo {author} {\bibfnamefont {S.}~\bibnamefont {Brierley}},\ }\bibfield
  {title} {\bibinfo {title} {Variational quantum computation of excited
  states},\ }\href {https://doi.org/10.22331/q-2019-07-01-156} {\bibfield
  {journal} {\bibinfo  {journal} {Quantum}\ }\textbf {\bibinfo {volume} {3}},\
  \bibinfo {pages} {156} (\bibinfo {year} {2019})},\ \Eprint
  {https://arxiv.org/abs/1805.08138} {arXiv:1805.08138} \BibitemShut {NoStop}%
\bibitem [{\citenamefont {Parlett}(1998)}]{Parlett1998}%
  \BibitemOpen
  \bibfield  {author} {\bibinfo {author} {\bibfnamefont {B.~N.}\ \bibnamefont
  {Parlett}},\ }\href {https://doi.org/10.1137/1.9781611971163} {\emph
  {\bibinfo {title} {The Symmetric Eigenvalue Problem}}},\ \bibinfo {series}
  {Classics in Applied Mathematics}, Vol.~\bibinfo {volume} {20}\ (\bibinfo
  {publisher} {SIAM},\ \bibinfo {year} {1998})\BibitemShut {NoStop}%
\bibitem [{\citenamefont {Jarvis}\ \emph {et~al.}(1997)\citenamefont {Jarvis},
  \citenamefont {White}, \citenamefont {Godby},\ and\ \citenamefont
  {Payne}}]{Jarvis1997}%
  \BibitemOpen
  \bibfield  {author} {\bibinfo {author} {\bibfnamefont {M.~R.}\ \bibnamefont
  {Jarvis}}, \bibinfo {author} {\bibfnamefont {I.~D.}\ \bibnamefont {White}},
  \bibinfo {author} {\bibfnamefont {R.~W.}\ \bibnamefont {Godby}},\ and\
  \bibinfo {author} {\bibfnamefont {M.~C.}\ \bibnamefont {Payne}},\ }\bibfield
  {title} {\bibinfo {title} {Supercell technique for total-energy calculations
  of finite charged and polar systems},\ }\href
  {https://doi.org/10.1103/PhysRevB.56.14972} {\bibfield  {journal} {\bibinfo
  {journal} {Phys. Rev. B}\ }\textbf {\bibinfo {volume} {56}},\ \bibinfo
  {pages} {14972} (\bibinfo {year} {1997})},\ \Eprint
  {https://arxiv.org/abs/cond-mat/9709234} {arXiv:cond-mat/9709234}
  \BibitemShut {NoStop}%
\bibitem [{\citenamefont {Rozzi}\ \emph {et~al.}(2006)\citenamefont {Rozzi},
  \citenamefont {Varsano}, \citenamefont {Marini}, \citenamefont {Gross},\ and\
  \citenamefont {Rubio}}]{Rozzi2006}%
  \BibitemOpen
  \bibfield  {author} {\bibinfo {author} {\bibfnamefont {C.~A.}\ \bibnamefont
  {Rozzi}}, \bibinfo {author} {\bibfnamefont {D.}~\bibnamefont {Varsano}},
  \bibinfo {author} {\bibfnamefont {A.}~\bibnamefont {Marini}}, \bibinfo
  {author} {\bibfnamefont {E.~K.~U.}\ \bibnamefont {Gross}},\ and\ \bibinfo
  {author} {\bibfnamefont {A.}~\bibnamefont {Rubio}},\ }\bibfield  {title}
  {\bibinfo {title} {Exact {C}oulomb cutoff technique for supercell
  calculations},\ }\href {https://doi.org/10.1103/PhysRevB.73.205119}
  {\bibfield  {journal} {\bibinfo  {journal} {Phys. Rev. B}\ }\textbf {\bibinfo
  {volume} {73}},\ \bibinfo {pages} {205119} (\bibinfo {year} {2006})},\
  \Eprint {https://arxiv.org/abs/cond-mat/0601031} {arXiv:cond-mat/0601031}
  \BibitemShut {NoStop}%
\bibitem [{\citenamefont {M{\"o}tt{\"o}nen}\ \emph {et~al.}(2005)\citenamefont
  {M{\"o}tt{\"o}nen}, \citenamefont {Vartiainen}, \citenamefont {Bergholm},\
  and\ \citenamefont {Salomaa}}]{Mottonen2005}%
  \BibitemOpen
  \bibfield  {author} {\bibinfo {author} {\bibfnamefont {M.}~\bibnamefont
  {M{\"o}tt{\"o}nen}}, \bibinfo {author} {\bibfnamefont {J.~J.}\ \bibnamefont
  {Vartiainen}}, \bibinfo {author} {\bibfnamefont {V.}~\bibnamefont
  {Bergholm}},\ and\ \bibinfo {author} {\bibfnamefont {M.~M.}\ \bibnamefont
  {Salomaa}},\ }\bibfield  {title} {\bibinfo {title} {Transformation of quantum
  states using uniformly controlled rotations},\ }\href
  {https://doi.org/10.26421/QIC5.6-5} {\bibfield  {journal} {\bibinfo
  {journal} {Quantum Inf. Comput.}\ }\textbf {\bibinfo {volume} {5}},\ \bibinfo
  {pages} {467} (\bibinfo {year} {2005})},\ \Eprint
  {https://arxiv.org/abs/quant-ph/0407010} {arXiv:quant-ph/0407010}
  \BibitemShut {NoStop}%
\bibitem [{\citenamefont {Knyazev}(2001)}]{Knyazev2001}%
  \BibitemOpen
  \bibfield  {author} {\bibinfo {author} {\bibfnamefont {A.~V.}\ \bibnamefont
  {Knyazev}},\ }\bibfield  {title} {\bibinfo {title} {Toward the optimal
  preconditioned eigensolver: Locally optimal block preconditioned conjugate
  gradient method},\ }\href {https://doi.org/10.1137/S1064827500366124}
  {\bibfield  {journal} {\bibinfo  {journal} {SIAM J. Sci. Comput.}\ }\textbf
  {\bibinfo {volume} {23}},\ \bibinfo {pages} {517} (\bibinfo {year}
  {2001})}\BibitemShut {NoStop}%
\bibitem [{\citenamefont {Davidson}(1975)}]{Davidson1975}%
  \BibitemOpen
  \bibfield  {author} {\bibinfo {author} {\bibfnamefont {E.~R.}\ \bibnamefont
  {Davidson}},\ }\bibfield  {title} {\bibinfo {title} {The iterative
  calculation of a few of the lowest eigenvalues and corresponding eigenvectors
  of large real-symmetric matrices},\ }\href
  {https://doi.org/10.1016/0021-9991(75)90065-0} {\bibfield  {journal}
  {\bibinfo  {journal} {J. Comput. Phys.}\ }\textbf {\bibinfo {volume} {17}},\
  \bibinfo {pages} {87} (\bibinfo {year} {1975})}\BibitemShut {NoStop}%
\bibitem [{\citenamefont {Anderson}(1965)}]{Anderson1965}%
  \BibitemOpen
  \bibfield  {author} {\bibinfo {author} {\bibfnamefont {D.~G.}\ \bibnamefont
  {Anderson}},\ }\bibfield  {title} {\bibinfo {title} {Iterative procedures for
  nonlinear integral equations},\ }\href
  {https://doi.org/10.1145/321296.321305} {\bibfield  {journal} {\bibinfo
  {journal} {J. ACM}\ }\textbf {\bibinfo {volume} {12}},\ \bibinfo {pages}
  {547} (\bibinfo {year} {1965})}\BibitemShut {NoStop}%
\bibitem [{\citenamefont {Mermin}(1965)}]{Mermin1965}%
  \BibitemOpen
  \bibfield  {author} {\bibinfo {author} {\bibfnamefont {N.~D.}\ \bibnamefont
  {Mermin}},\ }\bibfield  {title} {\bibinfo {title} {Thermal properties of the
  inhomogeneous electron gas},\ }\href
  {https://doi.org/10.1103/PhysRev.137.A1441} {\bibfield  {journal} {\bibinfo
  {journal} {Phys. Rev.}\ }\textbf {\bibinfo {volume} {137}},\ \bibinfo {pages}
  {A1441} (\bibinfo {year} {1965})}\BibitemShut {NoStop}%
\bibitem [{\citenamefont {Liang}\ and\ \citenamefont
  {Head-Gordon}(2004)}]{Liang2004}%
  \BibitemOpen
  \bibfield  {author} {\bibinfo {author} {\bibfnamefont {W.}~\bibnamefont
  {Liang}}\ and\ \bibinfo {author} {\bibfnamefont {M.}~\bibnamefont
  {Head-Gordon}},\ }\bibfield  {title} {\bibinfo {title} {Approaching the basis
  set limit in density functional theory calculations using dual basis sets
  without diagonalization},\ }\href {https://doi.org/10.1021/jp0374713}
  {\bibfield  {journal} {\bibinfo  {journal} {J. Phys. Chem. A}\ }\textbf
  {\bibinfo {volume} {108}},\ \bibinfo {pages} {3206} (\bibinfo {year}
  {2004})}\BibitemShut {NoStop}%
\bibitem [{\citenamefont {L{\"o}wdin}(1970)}]{Lowdin1970}%
  \BibitemOpen
  \bibfield  {author} {\bibinfo {author} {\bibfnamefont {P.-O.}\ \bibnamefont
  {L{\"o}wdin}},\ }\bibfield  {title} {\bibinfo {title} {On the
  nonorthogonality problem},\ }in\ \href
  {https://doi.org/10.1016/S0065-3276(08)60339-1} {\emph {\bibinfo {booktitle}
  {Advances in Quantum Chemistry}}},\ Vol.~\bibinfo {volume} {5}\ (\bibinfo
  {publisher} {Academic Press},\ \bibinfo {year} {1970})\ pp.\ \bibinfo {pages}
  {185--199}\BibitemShut {NoStop}%
\bibitem [{\citenamefont {Louie}\ \emph {et~al.}(1979)\citenamefont {Louie},
  \citenamefont {Ho},\ and\ \citenamefont {Cohen}}]{Louie1979}%
  \BibitemOpen
  \bibfield  {author} {\bibinfo {author} {\bibfnamefont {S.~G.}\ \bibnamefont
  {Louie}}, \bibinfo {author} {\bibfnamefont {K.-M.}\ \bibnamefont {Ho}},\ and\
  \bibinfo {author} {\bibfnamefont {M.~L.}\ \bibnamefont {Cohen}},\ }\bibfield
  {title} {\bibinfo {title} {Self-consistent mixed-basis approach to the
  electronic structure of solids},\ }\href
  {https://doi.org/10.1103/PhysRevB.19.1774} {\bibfield  {journal} {\bibinfo
  {journal} {Phys. Rev. B}\ }\textbf {\bibinfo {volume} {19}},\ \bibinfo
  {pages} {1774} (\bibinfo {year} {1979})}\BibitemShut {NoStop}%
\bibitem [{\citenamefont {Herring}(1940)}]{Herring1940}%
  \BibitemOpen
  \bibfield  {author} {\bibinfo {author} {\bibfnamefont {C.}~\bibnamefont
  {Herring}},\ }\bibfield  {title} {\bibinfo {title} {A new method for
  calculating wave functions in crystals},\ }\href
  {https://doi.org/10.1103/PhysRev.57.1169} {\bibfield  {journal} {\bibinfo
  {journal} {Phys. Rev.}\ }\textbf {\bibinfo {volume} {57}},\ \bibinfo {pages}
  {1169} (\bibinfo {year} {1940})}\BibitemShut {NoStop}%
\bibitem [{\citenamefont {Lippert}\ \emph {et~al.}(1997)\citenamefont
  {Lippert}, \citenamefont {Hutter},\ and\ \citenamefont
  {Parrinello}}]{Lippert1997}%
  \BibitemOpen
  \bibfield  {author} {\bibinfo {author} {\bibfnamefont {G.}~\bibnamefont
  {Lippert}}, \bibinfo {author} {\bibfnamefont {J.}~\bibnamefont {Hutter}},\
  and\ \bibinfo {author} {\bibfnamefont {M.}~\bibnamefont {Parrinello}},\
  }\bibfield  {title} {\bibinfo {title} {A hybrid {G}aussian and plane wave
  density functional scheme},\ }\href
  {https://doi.org/10.1080/00268979709482119} {\bibfield  {journal} {\bibinfo
  {journal} {Mol. Phys.}\ }\textbf {\bibinfo {volume} {92}},\ \bibinfo {pages}
  {477} (\bibinfo {year} {1997})}\BibitemShut {NoStop}%
\bibitem [{\citenamefont {Lippert}\ \emph {et~al.}(1999)\citenamefont
  {Lippert}, \citenamefont {Hutter},\ and\ \citenamefont
  {Parrinello}}]{Lippert1999}%
  \BibitemOpen
  \bibfield  {author} {\bibinfo {author} {\bibfnamefont {G.}~\bibnamefont
  {Lippert}}, \bibinfo {author} {\bibfnamefont {J.}~\bibnamefont {Hutter}},\
  and\ \bibinfo {author} {\bibfnamefont {M.}~\bibnamefont {Parrinello}},\
  }\bibfield  {title} {\bibinfo {title} {The {G}aussian and
  augmented-plane-wave density functional method for ab initio molecular
  dynamics simulations},\ }\href {https://doi.org/10.1007/s002140050523}
  {\bibfield  {journal} {\bibinfo  {journal} {Theor. Chem. Acc.}\ }\textbf
  {\bibinfo {volume} {103}},\ \bibinfo {pages} {124} (\bibinfo {year}
  {1999})}\BibitemShut {NoStop}%
\bibitem [{\citenamefont {Slater}(1937)}]{Slater1937}%
  \BibitemOpen
  \bibfield  {author} {\bibinfo {author} {\bibfnamefont {J.~C.}\ \bibnamefont
  {Slater}},\ }\bibfield  {title} {\bibinfo {title} {Wave functions in a
  periodic potential},\ }\href {https://doi.org/10.1103/PhysRev.51.846}
  {\bibfield  {journal} {\bibinfo  {journal} {Phys. Rev.}\ }\textbf {\bibinfo
  {volume} {51}},\ \bibinfo {pages} {846} (\bibinfo {year} {1937})}\BibitemShut
  {NoStop}%
\bibitem [{\citenamefont {Lanczos}(1950)}]{Lanczos1950}%
  \BibitemOpen
  \bibfield  {author} {\bibinfo {author} {\bibfnamefont {C.}~\bibnamefont
  {Lanczos}},\ }\bibfield  {title} {\bibinfo {title} {An iteration method for
  the solution of the eigenvalue problem of linear differential and integral
  operators},\ }\href {https://doi.org/10.6028/jres.045.026} {\bibfield
  {journal} {\bibinfo  {journal} {J. Res. Natl. Bur. Stand.}\ }\textbf
  {\bibinfo {volume} {45}},\ \bibinfo {pages} {255} (\bibinfo {year}
  {1950})}\BibitemShut {NoStop}%
\bibitem [{\citenamefont {Wecker}\ \emph {et~al.}(2015)\citenamefont {Wecker},
  \citenamefont {Hastings},\ and\ \citenamefont {Troyer}}]{Wecker2015}%
  \BibitemOpen
  \bibfield  {author} {\bibinfo {author} {\bibfnamefont {D.}~\bibnamefont
  {Wecker}}, \bibinfo {author} {\bibfnamefont {M.~B.}\ \bibnamefont
  {Hastings}},\ and\ \bibinfo {author} {\bibfnamefont {M.}~\bibnamefont
  {Troyer}},\ }\bibfield  {title} {\bibinfo {title} {Towards practical quantum
  variational algorithms},\ }\href {https://doi.org/10.1103/PhysRevA.92.042303}
  {\bibfield  {journal} {\bibinfo  {journal} {Phys. Rev. A}\ }\textbf {\bibinfo
  {volume} {92}},\ \bibinfo {pages} {042303} (\bibinfo {year} {2015})},\
  \Eprint {https://arxiv.org/abs/1507.08969} {arXiv:1507.08969} \BibitemShut
  {NoStop}%
\bibitem [{\citenamefont {Dunning}(1989)}]{Dunning1989}%
  \BibitemOpen
  \bibfield  {author} {\bibinfo {author} {\bibfnamefont {T.~H.}\ \bibnamefont
  {Dunning}, \bibfnamefont {Jr.}},\ }\bibfield  {title} {\bibinfo {title}
  {{G}aussian basis sets for use in correlated molecular calculations. {I}.
  {T}he atoms boron through neon and hydrogen},\ }\href
  {https://doi.org/10.1063/1.456153} {\bibfield  {journal} {\bibinfo  {journal}
  {J. Chem. Phys.}\ }\textbf {\bibinfo {volume} {90}},\ \bibinfo {pages} {1007}
  (\bibinfo {year} {1989})}\BibitemShut {NoStop}%
\bibitem [{\citenamefont {Sun}\ \emph {et~al.}(2018)\citenamefont {Sun},
  \citenamefont {Berkelbach}, \citenamefont {Blunt}, \citenamefont {Booth},
  \citenamefont {Guo}, \citenamefont {Li}, \citenamefont {Liu}, \citenamefont
  {McClain}, \citenamefont {Sayfutyarova}, \citenamefont {Sharma},
  \citenamefont {Wouters},\ and\ \citenamefont {Chan}}]{Sun2018}%
  \BibitemOpen
  \bibfield  {author} {\bibinfo {author} {\bibfnamefont {Q.}~\bibnamefont
  {Sun}}, \bibinfo {author} {\bibfnamefont {T.~C.}\ \bibnamefont {Berkelbach}},
  \bibinfo {author} {\bibfnamefont {N.~S.}\ \bibnamefont {Blunt}}, \bibinfo
  {author} {\bibfnamefont {G.~H.}\ \bibnamefont {Booth}}, \bibinfo {author}
  {\bibfnamefont {S.}~\bibnamefont {Guo}}, \bibinfo {author} {\bibfnamefont
  {Z.}~\bibnamefont {Li}}, \bibinfo {author} {\bibfnamefont {J.}~\bibnamefont
  {Liu}}, \bibinfo {author} {\bibfnamefont {J.~D.}\ \bibnamefont {McClain}},
  \bibinfo {author} {\bibfnamefont {E.~R.}\ \bibnamefont {Sayfutyarova}},
  \bibinfo {author} {\bibfnamefont {S.}~\bibnamefont {Sharma}}, \bibinfo
  {author} {\bibfnamefont {S.}~\bibnamefont {Wouters}},\ and\ \bibinfo {author}
  {\bibfnamefont {G.~K.-L.}\ \bibnamefont {Chan}},\ }\bibfield  {title}
  {\bibinfo {title} {{PySCF}: the {P}ython-based simulations of chemistry
  framework},\ }\href {https://doi.org/10.1002/wcms.1340} {\bibfield  {journal}
  {\bibinfo  {journal} {WIREs Computational Molecular Science}\ }\textbf
  {\bibinfo {volume} {8}},\ \bibinfo {pages} {e1340} (\bibinfo {year}
  {2018})}\BibitemShut {NoStop}%
\bibitem [{\citenamefont {Huggins}\ \emph {et~al.}(2021)\citenamefont
  {Huggins}, \citenamefont {McClean}, \citenamefont {Rubin}, \citenamefont
  {Jiang}, \citenamefont {Wiebe}, \citenamefont {Whaley},\ and\ \citenamefont
  {Babbush}}]{Huggins2021}%
  \BibitemOpen
  \bibfield  {author} {\bibinfo {author} {\bibfnamefont {W.~J.}\ \bibnamefont
  {Huggins}}, \bibinfo {author} {\bibfnamefont {J.~R.}\ \bibnamefont
  {McClean}}, \bibinfo {author} {\bibfnamefont {N.~C.}\ \bibnamefont {Rubin}},
  \bibinfo {author} {\bibfnamefont {Z.}~\bibnamefont {Jiang}}, \bibinfo
  {author} {\bibfnamefont {N.}~\bibnamefont {Wiebe}}, \bibinfo {author}
  {\bibfnamefont {K.~B.}\ \bibnamefont {Whaley}},\ and\ \bibinfo {author}
  {\bibfnamefont {R.}~\bibnamefont {Babbush}},\ }\bibfield  {title} {\bibinfo
  {title} {Efficient and noise resilient measurements for quantum chemistry on
  near-term quantum computers},\ }\href
  {https://doi.org/10.1038/s41534-020-00341-7} {\bibfield  {journal} {\bibinfo
  {journal} {npj Quantum Inf.}\ }\textbf {\bibinfo {volume} {7}},\ \bibinfo
  {pages} {23} (\bibinfo {year} {2021})},\ \Eprint
  {https://arxiv.org/abs/1907.13117} {arXiv:1907.13117} \BibitemShut {NoStop}%
\bibitem [{\citenamefont {Verteletskyi}\ \emph {et~al.}(2020)\citenamefont
  {Verteletskyi}, \citenamefont {Yen},\ and\ \citenamefont
  {Izmaylov}}]{Verteletskyi2020}%
  \BibitemOpen
  \bibfield  {author} {\bibinfo {author} {\bibfnamefont {V.}~\bibnamefont
  {Verteletskyi}}, \bibinfo {author} {\bibfnamefont {T.-C.}\ \bibnamefont
  {Yen}},\ and\ \bibinfo {author} {\bibfnamefont {A.~F.}\ \bibnamefont
  {Izmaylov}},\ }\bibfield  {title} {\bibinfo {title} {Measurement optimization
  in the variational quantum eigensolver using a minimum clique cover},\ }\href
  {https://doi.org/10.1063/1.5141458} {\bibfield  {journal} {\bibinfo
  {journal} {J. Chem. Phys.}\ }\textbf {\bibinfo {volume} {152}},\ \bibinfo
  {pages} {124114} (\bibinfo {year} {2020})},\ \Eprint
  {https://arxiv.org/abs/1907.03358} {arXiv:1907.03358} \BibitemShut {NoStop}%
\bibitem [{\citenamefont {Crawford}\ \emph {et~al.}(2021)\citenamefont
  {Crawford}, \citenamefont {van Straaten}, \citenamefont {Wang}, \citenamefont
  {Parks}, \citenamefont {Campbell},\ and\ \citenamefont
  {Brierley}}]{Crawford2021}%
  \BibitemOpen
  \bibfield  {author} {\bibinfo {author} {\bibfnamefont {O.}~\bibnamefont
  {Crawford}}, \bibinfo {author} {\bibfnamefont {B.}~\bibnamefont {van
  Straaten}}, \bibinfo {author} {\bibfnamefont {D.}~\bibnamefont {Wang}},
  \bibinfo {author} {\bibfnamefont {T.}~\bibnamefont {Parks}}, \bibinfo
  {author} {\bibfnamefont {E.}~\bibnamefont {Campbell}},\ and\ \bibinfo
  {author} {\bibfnamefont {S.}~\bibnamefont {Brierley}},\ }\bibfield  {title}
  {\bibinfo {title} {Efficient quantum measurement of {Pauli} operators in the
  presence of finite sampling error},\ }\href
  {https://doi.org/10.22331/q-2021-01-20-385} {\bibfield  {journal} {\bibinfo
  {journal} {Quantum}\ }\textbf {\bibinfo {volume} {5}},\ \bibinfo {pages}
  {385} (\bibinfo {year} {2021})},\ \Eprint {https://arxiv.org/abs/1908.06942}
  {arXiv:1908.06942} \BibitemShut {NoStop}%
\bibitem [{\citenamefont {{IonQ, Inc.}}(2026)}]{IonQForte}%
  \BibitemOpen
  \bibfield  {author} {\bibinfo {author} {\bibnamefont {{IonQ, Inc.}}},\
  }\href@noop {} {\bibinfo {title} {{IonQ Forte}}},\ \bibinfo {howpublished}
  {\url{https://ionq.com/quantum-systems/forte}} (\bibinfo {year} {2026}),\
  \bibinfo {note} {accessed 2026-09-27}\BibitemShut {NoStop}%
\bibitem [{\citenamefont {Chen}\ \emph {et~al.}(2024)\citenamefont {Chen},
  \citenamefont {Nielsen}, \citenamefont {Ebert}, \citenamefont {Inlek},
  \citenamefont {Wright}, \citenamefont {Chaplin}, \citenamefont {Maksymov},
  \citenamefont {P{\'a}ez}, \citenamefont {Poudel}, \citenamefont {Maunz},\
  and\ \citenamefont {Gamble}}]{Chen2024Forte}%
  \BibitemOpen
  \bibfield  {author} {\bibinfo {author} {\bibfnamefont {J.-S.}\ \bibnamefont
  {Chen}}, \bibinfo {author} {\bibfnamefont {E.}~\bibnamefont {Nielsen}},
  \bibinfo {author} {\bibfnamefont {M.}~\bibnamefont {Ebert}}, \bibinfo
  {author} {\bibfnamefont {V.}~\bibnamefont {Inlek}}, \bibinfo {author}
  {\bibfnamefont {K.}~\bibnamefont {Wright}}, \bibinfo {author} {\bibfnamefont
  {V.}~\bibnamefont {Chaplin}}, \bibinfo {author} {\bibfnamefont
  {A.}~\bibnamefont {Maksymov}}, \bibinfo {author} {\bibfnamefont
  {E.}~\bibnamefont {P{\'a}ez}}, \bibinfo {author} {\bibfnamefont
  {A.}~\bibnamefont {Poudel}}, \bibinfo {author} {\bibfnamefont
  {P.}~\bibnamefont {Maunz}},\ and\ \bibinfo {author} {\bibfnamefont
  {J.}~\bibnamefont {Gamble}},\ }\bibfield  {title} {\bibinfo {title}
  {Benchmarking a trapped-ion quantum computer with 30 qubits},\ }\href
  {https://doi.org/10.22331/q-2024-11-07-1516} {\bibfield  {journal} {\bibinfo
  {journal} {Quantum}\ }\textbf {\bibinfo {volume} {8}},\ \bibinfo {pages}
  {1516} (\bibinfo {year} {2024})},\ \Eprint {https://arxiv.org/abs/2308.05071}
  {arXiv:2308.05071} \BibitemShut {NoStop}%
\bibitem [{\citenamefont {Maksymov}\ \emph {et~al.}(2023)\citenamefont
  {Maksymov}, \citenamefont {Nguyen}, \citenamefont {Nam},\ and\ \citenamefont
  {Markov}}]{Maksymov2023}%
  \BibitemOpen
  \bibfield  {author} {\bibinfo {author} {\bibfnamefont {A.}~\bibnamefont
  {Maksymov}}, \bibinfo {author} {\bibfnamefont {J.}~\bibnamefont {Nguyen}},
  \bibinfo {author} {\bibfnamefont {Y.}~\bibnamefont {Nam}},\ and\ \bibinfo
  {author} {\bibfnamefont {I.}~\bibnamefont {Markov}},\ }\href@noop {}
  {\bibinfo {title} {Enhancing quantum computer performance via
  symmetrization}} (\bibinfo {year} {2023}),\ \Eprint
  {https://arxiv.org/abs/2301.07233} {arXiv:2301.07233 [quant-ph]} \BibitemShut
  {NoStop}%
\bibitem [{\citenamefont {Vovrosh}\ \emph {et~al.}(2021)\citenamefont
  {Vovrosh}, \citenamefont {Khosla}, \citenamefont {Greenaway}, \citenamefont
  {Self}, \citenamefont {Kim},\ and\ \citenamefont {Knolle}}]{Vovrosh2021}%
  \BibitemOpen
  \bibfield  {author} {\bibinfo {author} {\bibfnamefont {J.}~\bibnamefont
  {Vovrosh}}, \bibinfo {author} {\bibfnamefont {K.~E.}\ \bibnamefont {Khosla}},
  \bibinfo {author} {\bibfnamefont {S.}~\bibnamefont {Greenaway}}, \bibinfo
  {author} {\bibfnamefont {C.}~\bibnamefont {Self}}, \bibinfo {author}
  {\bibfnamefont {M.~S.}\ \bibnamefont {Kim}},\ and\ \bibinfo {author}
  {\bibfnamefont {J.}~\bibnamefont {Knolle}},\ }\bibfield  {title} {\bibinfo
  {title} {Simple mitigation of global depolarizing errors in quantum
  simulations},\ }\href {https://doi.org/10.1103/PhysRevE.104.035309}
  {\bibfield  {journal} {\bibinfo  {journal} {Phys. Rev. E}\ }\textbf {\bibinfo
  {volume} {104}},\ \bibinfo {pages} {035309} (\bibinfo {year} {2021})},\
  \Eprint {https://arxiv.org/abs/2101.01690} {arXiv:2101.01690} \BibitemShut
  {NoStop}%
\bibitem [{\citenamefont {Urbanek}\ \emph {et~al.}(2021)\citenamefont
  {Urbanek}, \citenamefont {Nachman}, \citenamefont {Pascuzzi}, \citenamefont
  {He}, \citenamefont {Bauer},\ and\ \citenamefont {de~Jong}}]{Urbanek2021}%
  \BibitemOpen
  \bibfield  {author} {\bibinfo {author} {\bibfnamefont {M.}~\bibnamefont
  {Urbanek}}, \bibinfo {author} {\bibfnamefont {B.}~\bibnamefont {Nachman}},
  \bibinfo {author} {\bibfnamefont {V.~R.}\ \bibnamefont {Pascuzzi}}, \bibinfo
  {author} {\bibfnamefont {A.}~\bibnamefont {He}}, \bibinfo {author}
  {\bibfnamefont {C.~W.}\ \bibnamefont {Bauer}},\ and\ \bibinfo {author}
  {\bibfnamefont {W.~A.}\ \bibnamefont {de~Jong}},\ }\bibfield  {title}
  {\bibinfo {title} {Mitigating depolarizing noise on quantum computers with
  noise-estimation circuits},\ }\href
  {https://doi.org/10.1103/PhysRevLett.127.270502} {\bibfield  {journal}
  {\bibinfo  {journal} {Phys. Rev. Lett.}\ }\textbf {\bibinfo {volume} {127}},\
  \bibinfo {pages} {270502} (\bibinfo {year} {2021})},\ \Eprint
  {https://arxiv.org/abs/2103.08591} {arXiv:2103.08591} \BibitemShut {NoStop}%
\bibitem [{\citenamefont {Temme}\ \emph {et~al.}(2017)\citenamefont {Temme},
  \citenamefont {Bravyi},\ and\ \citenamefont {Gambetta}}]{Temme2017}%
  \BibitemOpen
  \bibfield  {author} {\bibinfo {author} {\bibfnamefont {K.}~\bibnamefont
  {Temme}}, \bibinfo {author} {\bibfnamefont {S.}~\bibnamefont {Bravyi}},\ and\
  \bibinfo {author} {\bibfnamefont {J.~M.}\ \bibnamefont {Gambetta}},\
  }\bibfield  {title} {\bibinfo {title} {Error mitigation for short-depth
  quantum circuits},\ }\href {https://doi.org/10.1103/PhysRevLett.119.180509}
  {\bibfield  {journal} {\bibinfo  {journal} {Phys. Rev. Lett.}\ }\textbf
  {\bibinfo {volume} {119}},\ \bibinfo {pages} {180509} (\bibinfo {year}
  {2017})},\ \Eprint {https://arxiv.org/abs/1612.02058} {arXiv:1612.02058}
  \BibitemShut {NoStop}%
\bibitem [{\citenamefont {Li}\ and\ \citenamefont {Benjamin}(2017)}]{Li2017}%
  \BibitemOpen
  \bibfield  {author} {\bibinfo {author} {\bibfnamefont {Y.}~\bibnamefont
  {Li}}\ and\ \bibinfo {author} {\bibfnamefont {S.~C.}\ \bibnamefont
  {Benjamin}},\ }\bibfield  {title} {\bibinfo {title} {Efficient variational
  quantum simulator incorporating active error minimization},\ }\href
  {https://doi.org/10.1103/PhysRevX.7.021050} {\bibfield  {journal} {\bibinfo
  {journal} {Phys. Rev. X}\ }\textbf {\bibinfo {volume} {7}},\ \bibinfo {pages}
  {021050} (\bibinfo {year} {2017})},\ \Eprint
  {https://arxiv.org/abs/1611.09301} {arXiv:1611.09301} \BibitemShut {NoStop}%
\bibitem [{\citenamefont {Kandala}\ \emph {et~al.}(2017)\citenamefont
  {Kandala}, \citenamefont {Mezzacapo}, \citenamefont {Temme}, \citenamefont
  {Takita}, \citenamefont {Brink}, \citenamefont {Chow},\ and\ \citenamefont
  {Gambetta}}]{Kandala2017}%
  \BibitemOpen
  \bibfield  {author} {\bibinfo {author} {\bibfnamefont {A.}~\bibnamefont
  {Kandala}}, \bibinfo {author} {\bibfnamefont {A.}~\bibnamefont {Mezzacapo}},
  \bibinfo {author} {\bibfnamefont {K.}~\bibnamefont {Temme}}, \bibinfo
  {author} {\bibfnamefont {M.}~\bibnamefont {Takita}}, \bibinfo {author}
  {\bibfnamefont {M.}~\bibnamefont {Brink}}, \bibinfo {author} {\bibfnamefont
  {J.~M.}\ \bibnamefont {Chow}},\ and\ \bibinfo {author} {\bibfnamefont
  {J.~M.}\ \bibnamefont {Gambetta}},\ }\bibfield  {title} {\bibinfo {title}
  {Hardware-efficient variational quantum eigensolver for small molecules and
  quantum magnets},\ }\href {https://doi.org/10.1038/nature23879} {\bibfield
  {journal} {\bibinfo  {journal} {Nature}\ }\textbf {\bibinfo {volume} {549}},\
  \bibinfo {pages} {242} (\bibinfo {year} {2017})},\ \Eprint
  {https://arxiv.org/abs/1704.05018} {arXiv:1704.05018} \BibitemShut {NoStop}%
\bibitem [{\citenamefont {Lehtola}\ \emph {et~al.}(2018)\citenamefont
  {Lehtola}, \citenamefont {Steigemann}, \citenamefont {Oliveira},\ and\
  \citenamefont {Marques}}]{Lehtola2018}%
  \BibitemOpen
  \bibfield  {author} {\bibinfo {author} {\bibfnamefont {S.}~\bibnamefont
  {Lehtola}}, \bibinfo {author} {\bibfnamefont {C.}~\bibnamefont {Steigemann}},
  \bibinfo {author} {\bibfnamefont {M.~J.~T.}\ \bibnamefont {Oliveira}},\ and\
  \bibinfo {author} {\bibfnamefont {M.~A.~L.}\ \bibnamefont {Marques}},\
  }\bibfield  {title} {\bibinfo {title} {Recent developments in {libxc} --- {A}
  comprehensive library of functionals for density functional theory},\ }\href
  {https://doi.org/10.1016/j.softx.2017.11.002} {\bibfield  {journal} {\bibinfo
   {journal} {SoftwareX}\ }\textbf {\bibinfo {volume} {7}},\ \bibinfo {pages}
  {1} (\bibinfo {year} {2018})}\BibitemShut {NoStop}%
\bibitem [{\citenamefont {Sun}\ \emph {et~al.}(2020)\citenamefont {Sun},
  \citenamefont {Zhang}, \citenamefont {Banerjee}, \citenamefont {Bao},
  \citenamefont {Barbry}, \citenamefont {Blunt}, \citenamefont {others},\ and\
  \citenamefont {Chan}}]{Sun2020}%
  \BibitemOpen
  \bibfield  {author} {\bibinfo {author} {\bibfnamefont {Q.}~\bibnamefont
  {Sun}}, \bibinfo {author} {\bibfnamefont {X.}~\bibnamefont {Zhang}}, \bibinfo
  {author} {\bibfnamefont {S.}~\bibnamefont {Banerjee}}, \bibinfo {author}
  {\bibfnamefont {P.}~\bibnamefont {Bao}}, \bibinfo {author} {\bibfnamefont
  {M.}~\bibnamefont {Barbry}}, \bibinfo {author} {\bibfnamefont {N.~S.}\
  \bibnamefont {Blunt}}, \bibinfo {author} {\bibnamefont {others}},\ and\
  \bibinfo {author} {\bibfnamefont {G.~K.-L.}\ \bibnamefont {Chan}},\
  }\bibfield  {title} {\bibinfo {title} {Recent developments in the {PySCF}
  program package},\ }\href {https://doi.org/10.1063/5.0006074} {\bibfield
  {journal} {\bibinfo  {journal} {J. Chem. Phys.}\ }\textbf {\bibinfo {volume}
  {153}},\ \bibinfo {pages} {024109} (\bibinfo {year} {2020})}\BibitemShut
  {NoStop}%
\bibitem [{\citenamefont {Virtanen}\ \emph {et~al.}(2020)\citenamefont
  {Virtanen}, \citenamefont {Gommers}, \citenamefont {Oliphant}, \citenamefont
  {Haberland}, \citenamefont {Reddy}, \citenamefont {Cournapeau} \emph
  {et~al.}}]{Virtanen2020}%
  \BibitemOpen
  \bibfield  {author} {\bibinfo {author} {\bibfnamefont {P.}~\bibnamefont
  {Virtanen}}, \bibinfo {author} {\bibfnamefont {R.}~\bibnamefont {Gommers}},
  \bibinfo {author} {\bibfnamefont {T.~E.}\ \bibnamefont {Oliphant}}, \bibinfo
  {author} {\bibfnamefont {M.}~\bibnamefont {Haberland}}, \bibinfo {author}
  {\bibfnamefont {T.}~\bibnamefont {Reddy}}, \bibinfo {author} {\bibfnamefont
  {D.}~\bibnamefont {Cournapeau}}, \emph {et~al.},\ }\bibfield  {title}
  {\bibinfo {title} {{SciPy} 1.0: fundamental algorithms for scientific
  computing in {P}ython},\ }\href {https://doi.org/10.1038/s41592-019-0686-2}
  {\bibfield  {journal} {\bibinfo  {journal} {Nat. Methods}\ }\textbf {\bibinfo
  {volume} {17}},\ \bibinfo {pages} {261} (\bibinfo {year} {2020})}\BibitemShut
  {NoStop}%
\bibitem [{\citenamefont {Lehoucq}\ \emph {et~al.}(1998)\citenamefont
  {Lehoucq}, \citenamefont {Sorensen},\ and\ \citenamefont
  {Yang}}]{Lehoucq1998}%
  \BibitemOpen
  \bibfield  {author} {\bibinfo {author} {\bibfnamefont {R.~B.}\ \bibnamefont
  {Lehoucq}}, \bibinfo {author} {\bibfnamefont {D.~C.}\ \bibnamefont
  {Sorensen}},\ and\ \bibinfo {author} {\bibfnamefont {C.}~\bibnamefont
  {Yang}},\ }\href {https://doi.org/10.1137/1.9780898719628} {\emph {\bibinfo
  {title} {{ARPACK} Users' Guide: Solution of Large-Scale Eigenvalue Problems
  with Implicitly Restarted {A}rnoldi Methods}}}\ (\bibinfo  {publisher}
  {SIAM},\ \bibinfo {address} {Philadelphia},\ \bibinfo {year}
  {1998})\BibitemShut {NoStop}%
\bibitem [{\citenamefont {Lebedev}\ and\ \citenamefont
  {Laikov}(1999)}]{Lebedev1999}%
  \BibitemOpen
  \bibfield  {author} {\bibinfo {author} {\bibfnamefont {V.~I.}\ \bibnamefont
  {Lebedev}}\ and\ \bibinfo {author} {\bibfnamefont {D.~N.}\ \bibnamefont
  {Laikov}},\ }\bibfield  {title} {\bibinfo {title} {A quadrature formula for
  the sphere of the 131st algebraic order of accuracy},\ }\href@noop {}
  {\bibfield  {journal} {\bibinfo  {journal} {Dokl. Math.}\ }\textbf {\bibinfo
  {volume} {59}},\ \bibinfo {pages} {477} (\bibinfo {year} {1999})}\BibitemShut
  {NoStop}%
\bibitem [{\citenamefont {Becke}(1988)}]{Becke1988}%
  \BibitemOpen
  \bibfield  {author} {\bibinfo {author} {\bibfnamefont {A.~D.}\ \bibnamefont
  {Becke}},\ }\bibfield  {title} {\bibinfo {title} {A multicenter numerical
  integration scheme for polyatomic molecules},\ }\href
  {https://doi.org/10.1063/1.454033} {\bibfield  {journal} {\bibinfo  {journal}
  {J. Chem. Phys.}\ }\textbf {\bibinfo {volume} {88}},\ \bibinfo {pages} {2547}
  (\bibinfo {year} {1988})}\BibitemShut {NoStop}%
\bibitem [{\citenamefont {Harris}\ \emph {et~al.}(2020)\citenamefont {Harris},
  \citenamefont {Millman}, \citenamefont {van~der Walt}, \citenamefont
  {Gommers}, \citenamefont {Virtanen}, \citenamefont {Cournapeau} \emph
  {et~al.}}]{Harris2020}%
  \BibitemOpen
  \bibfield  {author} {\bibinfo {author} {\bibfnamefont {C.~R.}\ \bibnamefont
  {Harris}}, \bibinfo {author} {\bibfnamefont {K.~J.}\ \bibnamefont {Millman}},
  \bibinfo {author} {\bibfnamefont {S.~J.}\ \bibnamefont {van~der Walt}},
  \bibinfo {author} {\bibfnamefont {R.}~\bibnamefont {Gommers}}, \bibinfo
  {author} {\bibfnamefont {P.}~\bibnamefont {Virtanen}}, \bibinfo {author}
  {\bibfnamefont {D.}~\bibnamefont {Cournapeau}}, \emph {et~al.},\ }\bibfield
  {title} {\bibinfo {title} {Array programming with {NumPy}},\ }\href
  {https://doi.org/10.1038/s41586-020-2649-2} {\bibfield  {journal} {\bibinfo
  {journal} {Nature}\ }\textbf {\bibinfo {volume} {585}},\ \bibinfo {pages}
  {357} (\bibinfo {year} {2020})}\BibitemShut {NoStop}%
\bibitem [{\citenamefont {Javadi-Abhari}\ \emph {et~al.}(2024)\citenamefont
  {Javadi-Abhari}, \citenamefont {Treinish}, \citenamefont {Krsulich},
  \citenamefont {Wood}, \citenamefont {Lishman}, \citenamefont {Gacon},
  \citenamefont {Martiel}, \citenamefont {Nation}, \citenamefont {Bishop},
  \citenamefont {Cross}, \citenamefont {Johnson},\ and\ \citenamefont
  {Gambetta}}]{Qiskit2024}%
  \BibitemOpen
  \bibfield  {author} {\bibinfo {author} {\bibfnamefont {A.}~\bibnamefont
  {Javadi-Abhari}}, \bibinfo {author} {\bibfnamefont {M.}~\bibnamefont
  {Treinish}}, \bibinfo {author} {\bibfnamefont {K.}~\bibnamefont {Krsulich}},
  \bibinfo {author} {\bibfnamefont {C.~J.}\ \bibnamefont {Wood}}, \bibinfo
  {author} {\bibfnamefont {J.}~\bibnamefont {Lishman}}, \bibinfo {author}
  {\bibfnamefont {J.}~\bibnamefont {Gacon}}, \bibinfo {author} {\bibfnamefont
  {S.}~\bibnamefont {Martiel}}, \bibinfo {author} {\bibfnamefont {P.~D.}\
  \bibnamefont {Nation}}, \bibinfo {author} {\bibfnamefont {L.~S.}\
  \bibnamefont {Bishop}}, \bibinfo {author} {\bibfnamefont {A.~W.}\
  \bibnamefont {Cross}}, \bibinfo {author} {\bibfnamefont {B.~R.}\ \bibnamefont
  {Johnson}},\ and\ \bibinfo {author} {\bibfnamefont {J.~M.}\ \bibnamefont
  {Gambetta}},\ }\href {https://doi.org/10.48550/arXiv.2405.08810} {\bibinfo
  {title} {Quantum computing with {Q}iskit}} (\bibinfo {year} {2024}),\ \Eprint
  {https://arxiv.org/abs/2405.08810} {arXiv:2405.08810 [quant-ph]} \BibitemShut
  {NoStop}%
\bibitem [{\citenamefont {Iten}\ \emph {et~al.}(2016)\citenamefont {Iten},
  \citenamefont {Colbeck}, \citenamefont {Kukuljan}, \citenamefont {Home},\
  and\ \citenamefont {Christandl}}]{Iten2016}%
  \BibitemOpen
  \bibfield  {author} {\bibinfo {author} {\bibfnamefont {R.}~\bibnamefont
  {Iten}}, \bibinfo {author} {\bibfnamefont {R.}~\bibnamefont {Colbeck}},
  \bibinfo {author} {\bibfnamefont {I.}~\bibnamefont {Kukuljan}}, \bibinfo
  {author} {\bibfnamefont {J.}~\bibnamefont {Home}},\ and\ \bibinfo {author}
  {\bibfnamefont {M.}~\bibnamefont {Christandl}},\ }\bibfield  {title}
  {\bibinfo {title} {Quantum circuits for isometries},\ }\href
  {https://doi.org/10.1103/PhysRevA.93.032318} {\bibfield  {journal} {\bibinfo
  {journal} {Phys. Rev. A}\ }\textbf {\bibinfo {volume} {93}},\ \bibinfo
  {pages} {032318} (\bibinfo {year} {2016})},\ \Eprint
  {https://arxiv.org/abs/1501.06911} {arXiv:1501.06911} \BibitemShut {NoStop}%
\bibitem [{\citenamefont {{IonQ, Inc.}}(2025)}]{QiskitIonQ}%
  \BibitemOpen
  \bibfield  {author} {\bibinfo {author} {\bibnamefont {{IonQ, Inc.}}},\
  }\href@noop {} {\bibinfo {title} {qiskit-ionq: {IonQ} provider for
  {Q}iskit}},\ \bibinfo {howpublished}
  {\url{https://github.com/qiskit-community/qiskit-ionq}} (\bibinfo {year}
  {2025}),\ \bibinfo {note} {version 1.0.2}\BibitemShut {NoStop}%
\end{thebibliography}%

\end{document}